\documentclass[12pt]{article}
\usepackage{latexsym}
\usepackage{amsmath,amsfonts}
\usepackage{times}
\allowdisplaybreaks[4]

\catcode`\@=11

\newcount\hour
\newcount\minute
\newtoks\amorpm \hour=\time\divide\hour by 60\minute
=\time{\multiply\hour by 60 \global\advance\minute by-\hour}
\edef\standardtime{{\ifnum\hour<12 \global\amorpm={am}%
        \else\global\amorpm={pm}\advance\hour by-12 \fi
        \ifnum\hour=0 \hour=12 \fi
        \number\hour:\ifnum\minute<10
        0\fi\number\minute\the\amorpm}}
\edef\militarytime{\number\hour:\ifnum\minute<10
0\fi\number\minute}

\def\draftlabel#1{{\@bsphack\if@filesw {\let\thepage\relax
   \xdef\@gtempa{\write\@auxout{\string
      \newlabel{#1}{{\@currentlabel}{\thepage}}}}}\@gtempa
   \if@nobreak \ifvmode\nobreak\fi\fi\fi\@esphack}
        \gdef\@eqnlabel{#1}}
\def\@eqnlabel{}
\def\@vacuum{}
\def\marginnote#1{}
\def\draftmarginnote#1{\marginpar{\raggedright\scriptsize\tt#1}}
\def\draft{
        \pagestyle{plain}
        \overfullrule=2pt
        \oddsidemargin -.5truein
        \def\@oddhead{\sl \phantom{\today\quad\militarytime} \hfil
        \smash{\Large\sl DRAFT} \hfil \today\quad\militarytime}
        \let\@evenhead\@oddhead
        \let\label=\draftlabel
        \let\marginnote=\draftmarginnote
        \def\ps@empty{\let\@mkboth\@gobbletwo
        \def\@oddfoot{\hfil \smash{\Large\sl DRAFT} \hfil}
        \let\@evenfoot\@oddhead}
        \def\@eqnnum{(\theequation)\rlap{\kern\marginparsep\tt\@eqnlabel}%
        \global\let\@eqnlabel\@vacuum}  }

\newcommand{\rf}[1]{(\ref{#1})}
\renewcommand{\theequation}{\thesection.\arabic{equation}}
\renewcommand{\thefootnote}{\fnsymbol{footnote}}
\newcommand{\newsection}{    
\setcounter{equation}{0}\section}
\def\appendix#1{\addtocounter{section}{1}\setcounter{equation}{0}
\renewcommand{\thesection}{\Alph{section}}
\section*{Appendix \thesection\protect\indent \parbox[t]{11.15cm}{#1}}
\addcontentsline{toc}{section}{Appendix \thesection\ \ \ #1}}

\def\MM{{\cal M}}
\def\LL{{\cal L}}
\def\TT{{\cal T}}

\def\half{\frac{1}{2}}

\def\hbf{{\bf h}}

\def\ibf{{\bf i}}
\def\iibf{{\bf ii}}
\def\iiibf{{\bf iii}}
\def\ivbf{{\bf iv}}
\def\vbf{{\bf v}}
\def\vibf{{\bf vi}}

\newcommand{\Co}{\mathbb{C}}

\def\be{\begin{equation}}
\def\ee{\end{equation}}
\def\beq{\begin{eqnarray}}
\def\eeq{\end{eqnarray}}

\def\phik{|\phi\rangle}
\def\phibr{\langle\phi|}
\def\psik{|\psi\rangle}
\def\psibr{\langle\psi|}
\def\half{{\frac{1}{2}}}

\def\intrm{{\rm int}}
\def\irm{{\rm i}}
\def\ub{\bar{u}}
\def\vb{\bar{v}}
\def\field{{\rm field}}
\def\ferm{{\rm ferm}}
\def\diff{{\rm diff}}

\def\oplussm{{\scriptscriptstyle \oplus}}
\def\ominussm{{\scriptscriptstyle \ominus}}

\def\cur{{\rm cur}}
\def\sh{{\rm sh}}
\def\eff{{\rm eff}}

\def\AdSsm{{\scriptscriptstyle AdS}}
\def\CFTsm{{\scriptscriptstyle CFT}}
\def\Rsm{{\scriptscriptstyle R}}
\def\Lsm{{\scriptscriptstyle L}}

\begin{document}


\begin{flushright}
FIAN-TD-2014-16 \hspace{1.2cm} {}~\\
arXiv: yymm.nnnn [hep-th]
\end{flushright}

\vspace{1cm}

\begin{center}

{\Large \bf Mixed-symmetry fields in AdS(5), conformal fields,

\medskip
and AdS/CFT}

\vspace{2.5cm}

R.R. Metsaev\footnote{ E-mail: metsaev@lpi.ru }

\vspace{1cm}

{\it Department of Theoretical Physics, P.N. Lebedev Physical
Institute, \\ Leninsky prospect 53,  Moscow 119991, Russia }

\vspace{3.5cm}

{\bf Abstract}

\end{center}

Mixed-symmetry arbitrary spin massive, massless, and self-dual massive fields in AdS(5) are studied. Light-cone gauge actions for such fields leading to decoupled equations of motion are constructed. Light-cone gauge formulation of mixed-symmetry anomalous conformal currents and shadows in 4d flat space is also developed. AdS/CFT correspondence  for normalizable and non-normalizable modes of mixed-symmetry AdS fields and the respective boundary mixed-symmetry anomalous conformal currents and shadows is studied. We demonstrate that the light-cone gauge action for massive mixed-symmetry AdS field evaluated on solution of the Dirichlet problem amounts to the light-cone gauge 2-point vertex of mixed-symmetry anomalous shadow. Also we show that UV divergence of the action for mixed-symmetry massive AdS field with some particular value of mass parameter evaluated on the  Dirichlet problem amounts to the action of long mixed-symmetry conformal field,
while UV divergence of the action for mixed-symmetry massless AdS field evaluated on the Dirichlet problem amounts to the action of short mixed-symmetry conformal field.
We speculate on string theory interpretation of a model which involves short low-spin conformal fields and long higher-spin conformal fields.

\newpage
\renewcommand{\thefootnote}{\arabic{footnote}}
\setcounter{footnote}{0}

\section{Introduction}

Conjectured string/gauge theory duality in Ref.\cite{malda} and light-cone gauge approach to string theory in AdS space in Ref.\cite{Metsaev:2000yf} triggered our interest to light-cone gauge formulation of field dynamics in AdS space. This is to say that the light-cone approach to string theory in AdS space implies the corresponding light-cone formulation for target space fields in AdS. It is expected that arbitrary spin totally symmetric and mixed-symmetry massive AdS fields together with low-spin massless fields form the spectrum of states of AdS string. Therefore a better understanding of the light-cone formulation for arbitrary spin massive AdS fields might be useful for discussion of
various aspects of string theory in AdS space. This is long-term motivation of our research  in this paper.

Before describing the particular problem we investigate in this paper, let us discuss group theoretical interpretation of fields in $AdS_5$ we are going to study. Massive and massless fields in $AdS_5$ space are associated with
unitary lowest weight representations of the
$so(4,2)$ algebra. A unitary positive-energy lowest weight irreducible
representation of the $so(4,2)$ algebra is denoted by $D(E_0,\hbf)$, where a label $E_0$ stands for the lowest eigenvalue of the energy
operator, while a label $\hbf = (h_1, h_2)$ is the highest
weight for representation of the $so(4)$ algebra. For bosonic fields, the
highest weights $h_1$, $h_2$ are integers, while for fermionic fields, the highest weight $h_1$, $h_2$ are  half-integers. These weights satisfy the well known restriction
\be
h_1 \geq |h_2|\,.
\ee
We recall that the field with $\hbf =(0,0)$ is a scalar bosonic field, while the field with $\hbf =(1/2,\pm1/2)$ is a spin one-half fermionic field.%
\footnote{
For the reader's convenience, we note that the field with $\hbf=(1,0)$ is spin-1
field, while the field with $\hbf = (2,0)$ is a spin-2
field. Sometimes, the labels $h_i$ are referred to as Gelfand--Zeitlin labels.
The Gelfand-Zeitlin labels are connected with the Dynkin labels $h_i^D$ by the relations
$(h^D_1,h^D_2)= (h_1-h_2, h_1 + h_2)$.}
The labels $E_0$ and $\hbf$ associated with the remaining fields in
$AdS_5$ satisfy the following restrictions \cite{Mack:1975je},%
\footnote{ Generalization of restrictions \rf{13102014-15}-\rf{13102014-19} to the case of $so(d,2)$ algebra with arbitrary $d$ may be found in Ref.\cite{Metsaev:1995re}.}
\beq
\label{13102014-15} && \hspace{-0.8cm} E_0 > h_1 + 1\,, \qquad h_1 = |h_2|>1/2\,, \hspace{2.5cm} \hbox{ self-dual massive fields}
\\
\label{13102014-16} && \hspace{-0.8cm} E_0 = h_1 + 2\,, \qquad h_1 > |h_2|>1/2\,,  \hspace{2.5cm} \hbox{ mixed-symmetry massless fields}
\\
\label{13102014-17} && \hspace{-0.8cm} E_0 > h_1 + 2\,, \qquad h_1 > |h_2|>1/2\,, \hspace{2.5cm} \hbox{ mixed-symmetry massive fields}
\\
\label{13102014-18} && \hspace{-0.8cm} E_0 = h_1 + 2\,, \qquad h_1 > |h_2|,\quad h_2 = 0, \pm 1/2\,,  \qquad \hbox{ totally symmetric massless fields}\qquad
\\
\label{13102014-19} && \hspace{-0.8cm} E_0 > h_1 + 2\,, \qquad h_1 > |h_2|, \quad h_2=0,\pm 1/2\,, \qquad \hbox{ totally symmetric massive fields}\qquad
\eeq
Often, in place of the labels $E_0$, $h_1$, $h_2$, we use the respective labels $\kappa$, $j_1$, $j_2$ defined by the relations
\be \label{13102014-18-x1}
\kappa \equiv E_0 - 2\,, \qquad j_1 \equiv \half(h_1 + h_2)\,, \qquad j_2 \equiv \half (h_1-h_2)\,.
\ee

For fields in \rf{13102014-18},\rf{13102014-19}, light-cone gauge formulation was developed in Refs.\cite{Metsaev:1999ui,Metsaev:2003cu}.%
\footnote{ In Refs.\cite{Metsaev:1999ui,Metsaev:2003cu}, light-cone gauge formulation was developed for totally symmetric fields in $AdS_{d+1}$ with arbitrary $d$, $d\geq 3$.
Light-cone gauge formulation for fields in $AdS_3$ was investigated in \cite{Metsaev:2000qb}.}
Light-cone gauge actions in Refs.\cite{Metsaev:1999ui,Metsaev:2003cu}
lead to {\it coupled} equations of motion for fields in \rf{13102014-18},\rf{13102014-19}. Light-cone gauge actions leading to {\it decoupled} equations of motion for fields in \rf{13102014-18},\rf{13102014-19} were obtained recently in Ref.\cite{Metsaev:2013kaa}.
For fields in \rf{13102014-15},\rf{13102014-16}, light-cone gauge formulation was developed in Ref.\cite{Metsaev:2002vr}, while, for fields in \rf{13102014-17}, light-cone gauge formulation was studied in Ref.\cite{Metsaev:2004ee}.%
\footnote{ We note that it is light-cone gauge approach in Refs.\cite{Metsaev:2002vr,Metsaev:2004ee} that made it possible to develop a Lagrangian formulation of AdS fields in \rf{13102014-15}-\rf{13102014-17} for the first time.
Also, for the reader's convenience, we recall that, in Lorentz covariant approaches, the mixed-symmetry fields in \rf{13102014-15}-\rf{13102014-17} are
described by tensor (or tensor-spinor) fields whose $so(4,1)$ space-time tensor indices have the structure of the Young tableaux with two rows.
Namely, mixed-symmetry bosonic fields are associated with Young tableaux having row lengths $h_1$, $|h_2|$, while mixed-symmetry fermionic fields are associated with
Young tableaux having row lengths $h_1-\half$, $|h_2|-
\half$.
}
For fields in \rf{13102014-15},\rf{13102014-16}, light-cone gauge actions in Ref.\cite{Metsaev:2002vr} lead to {\it decoupled} equations of motion, while, for fields in \rf{13102014-17}, light-cone gauge actions in Ref.\cite{Metsaev:2004ee} lead to {\it coupled} equations of motion.

Two main aims of this paper are as follows. The first aim is to develop light-cone gauge formulation which leads to {\it decoupled} equations of motion for mixed-symmetry massive fields \rf{13102014-17}. The second aim is to apply such light-cone formulation to study of AdS/CFT correspondence. In this paper, our primary interest are the massive mixed-symmetry fields.  The massless and self-dual massive fields are realized as appropriate limits of the massive fields.  This allows us to study a Lagrangian formulation and the AdS/CFT correspondence for the mixed-symmetry massive, mixed-symmetry massless, and self-dual massive fields on an equal footing.

Before proceeding to main theme of this paper let us briefly review various Lorentz covariant approaches in the literature which have been developed for the description of mixed-symmetry fields in AdS space. At the level of equations of motion given in Lorentz gauge, massless and massive mixed-symmetry fields in $AdS_{d+1}$, with arbitrary $d$ were studied in Refs.\cite{Metsaev:1995re,Metsaev:2003cu,Metsaev:1998xg}. On-shell BRST invariant formulation of massless and massive mixed-symmetry fields may be found in Ref.\cite{Alkalaev:2009vm}. On-shell formulation of massless mixed-symmetry fields in terms of generalized connections of AdS symmetry algebras is studied in Refs.\cite{Skvortsov:2009zu}. Lagrangian frame-like description of massless and massive mixed-symmetry fields may be found in the respective Refs.\cite{Alkalaev:2005kw} and Refs.\cite{Zinoviev:2009gh}. Lagrangian metric-like formulation for some particular mixed-symmetry massless field is considered in Ref.\cite{Brink:2000ag}. Important aspects of mixed-symmetry massless fields were studied in Refs.\cite{Boulanger:2008up}.  Maxwell-like Lagrangians for massless mixed-symmetry fields are considered in Ref.\cite{Campoleoni:2012th}. Discussion of Lagrangian formulation of massive fields corresponding to two-column Young tableaux may be found in Ref.\cite{deMedeiros:2003px}.  BFV-BRST Lagrangian formulation of mixed-symmetry massive fields is discussed in Ref.\cite{Reshetnyak:2010ga}. In the framework of world-line spinning particle approach, the self-dual massive fields are studied in Ref.\cite{Bastianelli:2014lia}. Interesting recent discussion of mixed-symmetry fields may be found in Ref.\cite{Vasiliev:2012tv}. Interacting mixed-symmetry fields were studied in Refs.\cite{Alkalaev:2010af}.%
\footnote{ Interacting mixed-symmetry fields in flat space were studied by using light-cone gauge approach in Refs.\cite{Metsaev:1993mj}-\cite{Metsaev:2007rn} and by BRST method in Refs.\cite{Koh:1986vg,Metsaev:2012uy} (see also Ref.\cite{Henneaux:2012wg}). Taking into account successful applications of these two approaches in string theory (see, e.g., Ref.\cite{Green:1983hw}), the light-cone gauge and BRST approaches seem to be promising for developing Lagrangian formulation of interacting mixed-symmetry AdS fields.}

This paper is organized as follows. In Sec.\ref{mix-lagr}, we
develop light-cone gauge Lagrangian formulation which leads to decoupled equations of motion for mixed-symmetry massive fields in $AdS_5$. Also we
describe light-cone gauge realization of relativistic symmetries of actions for such fields. In Secs.\ref{masless},\ref{self-dual}, mixed-symmetry massless and self-dual massive fields are considered.  We demonstrate how light-cone gauge Lagrangian for such fields can be obtained from the Lagrangian for the mixed-symmetry massive field. In Sec.\ref{mix-cur-sh}, light-cone gauge formulation for mixed-symmetry anomalous conformal currents and shadows is developed. Two-point vertices for such currents and shadows are discussed. Section \ref{ads-cft} is devoted to study of the AdS/CFT correspondence for mixed-symmetry AdS fields and boundary mixed-symmetry conformal currents and shadows. Mixed-symmetry conformal fields are considered in Sec.\ref{conf-field}. We demonstrate how the action for such fields can be obtained from 2-point vertex of mixed-symmetry shadows.
Our conventions and notation are explained in Appendix A.

\newsection{ Mixed-symmetry massive fields in $AdS_5$ }\label{mix-lagr}

{\bf Field content}. To develop a light-cone gauge formulation of bosonic mixed-symmetry massive fields in $AdS_5$ we introduce the following set of complex-valued fields
\be \label{12102014-01}
\phi_{m_1,m_2}\,, \qquad m_1=-j_1,-j_1+1,\ldots, j_1\,, \qquad m_2=-j_2,-j_2+1,\ldots, j_2\,. %
\ee
In order to obtain the light-cone gauge description in an easy--to--use form,
we introduce oscillators $u_\tau$, $v_\tau$, $\tau=1,2$ and collect fields \rf{12102014-01} into ket-vector $|\phi\rangle$ defined by the relation
\be \label{12102014-02}
\phik  = \sum_{m_1=-j_1}^{j_1}\sum_{m_2=-j_2}^{j_2} \frac{u_1^{j_1+m_1} v_1^{j_1-m_1} u_2^{j_2+m_2} v_2^{j_2-m_2}}{\sqrt{(j_1+m_1)!(j_1-m_1)!(j_2+m_2)!(j_2-m_2)!}}\, \phi_{m_1,m_2}|0\rangle\,.
\ee
Commutation relations for the oscillators, hermitian conjugation rules, and the vacuum $|0\rangle$ are defined by the relations
\beq
\label{26102014-01} && [\ub_\tau, u_\sigma]=\delta_{\tau\sigma}, \qquad  [\vb_\tau, v_\sigma]=\delta_{\tau\sigma}\,, \qquad \tau,\sigma=1,2\,,
\\
\label{26102014-02} && \ub_\tau|0\rangle =  0\,, \qquad  \vb_\tau|0\rangle =  0\,, \qquad u_\tau^\dagger = \ub_\tau\,,\qquad  v_\tau^\dagger = \vb_\tau\,,
\eeq
where $\delta_{11} = \delta_{22}=1$, $\delta_{12}=\delta_{21}=0$.

\noindent {\bf Light-cone gauge action}. To discuss light-cone gauge action we use Poincar\'e parametrization of $AdS_5$ space given by (details of our notation may be found in Appendix A)
\be \label{12102014-03}
ds^2 = \frac{1}{z^2}(dx^a dx^a + dz dz)\,.
\ee
General representation for light-cone gauge action and Lagrangian found in Ref.\cite{Metsaev:1999ui} takes the form
\beq
\label{12102014-04} && S = \int  d^5 x \, \LL\,, \qquad d^5 x \equiv dx^+dx^- dx^1dx^2 dz\,,
\\
\label{12102014-05} && \LL =  \langle \phi|\bigl(\Box + \partial_z^2 -\frac{1}{z^2}A\bigr)|\phi\rangle\,,
\eeq
where, in light-cone frame, the D'Alembertian operator $\Box$ in $R^{3,1}$ is given by
\be \label{12102014-06}
\Box = 2\partial^+\partial^- + \partial^i \partial^i\,,\qquad i=1,2\,.
\ee
Bra-vector $\phibr$ in \rf{12102014-05} is defined as $\phibr\equiv (\phik)^\dagger$.
The operator $A$ appearing in \rf{12102014-05} is independent of space-time coordinates and their derivatives. This operator depends only on the oscillators. We refer
to the operator $A$ as $AdS$ mass operator. From \rf{12102014-05}, we see that all that is required to fix the Lagrangian is to find the operator $A$. Solution for the operator $A$ we found is given by
\beq
\label{12102014-07} A & = &  \nu^2- \frac{1}{4}\,,
\\
\label{12102014-08} && \nu = \kappa+ S_1 - S_2\,,\qquad \kappa \equiv E_0 -2\,,
\\
\label{12102014-09} && S_1 = \half (N_{u_1} - N_{v_1})\,, \qquad S_2 = \half (N_{u_2} - N_{v_2})\,,
\\
\label{12102014-11} && N_{u_\tau} \equiv u_\tau \ub_\tau\,, \qquad N_{v_\tau} \equiv v_\tau\vb_\tau\,, \qquad \tau =1,2\,.
\eeq
In terms of fields \rf{12102014-01}, Lagrangian \rf{12102014-05} takes the form
\be \label{12102014-12}
\LL  =   \sum_{m_1=-j_1}^{j_1} \sum_{m_2=-j_2}^{j_2} \phi_{m_1,m_2}^\dagger \left( \Box + \partial_z^2 - \frac{1}{z^2}\Bigl((\kappa + m_1-m_2)^2 -\frac{1}{4}\Bigr)\right) \phi_{m_1,m_2}\,.
\ee

From \rf{12102014-12}, we see that our Lagrangian leads to {\it decoupled} equations of motion for the fields $\phi_{m_1,m_2}$,
\be \label{12102014-14}
\left( \Box + \partial_z^2 - \frac{1}{z^2}\Bigl((\kappa + m_1-m_2)^2 -\frac{1}{4}\Bigr)\right) \phi_{m_1,m_2} = 0\,.
\ee

The following remarks are in order.

\noindent \ibf) Using field transformation rules given below (see formulas \rf{13102014-01d}, \rf{13102014-02d}), we verify that if a set of fields \rf{12102014-01} transform in the representation of the $so(4,2)$ algebra labeled by $\kappa,j_1,j_2$, then a set of the hermitian conjugated fields $\phi_{-m_2,-m_1}^\dagger$ with
$m_1$ and $m_2$ as in \rf{12102014-01} transform in the representation of the $so(4,2)$ algebra labeled by $\kappa, j_2,j_1$. From \rf{12102014-12}, we see that our Lagrangian involves the fields $\phi_{m_1,m_2}$ and their hermitian conjugated $\phi_{m_1,m_2}^\dagger$. This implies that our Lagrangian \rf{12102014-12} describes dynamics of fields associated with a direct sum of two representations of the $so(4,2)$ algebra which are labeled by $\kappa, j_1,j_2$ and $\kappa, j_2,j_1$.

\noindent \iibf) Lagrangian \rf{12102014-12} gives light-cone gauge description of the massive mixed-symmetry field. This Lagrangian can also be used for the description of the totally symmetric massive field. To get Lagrangian for the totally symmetric fields we set $h_2=0$ in \rf{12102014-12}, .i.e., $j_1=j_2$. We can impose then the reality condition on the fields in \rf{12102014-01}, $\phi_{-m_2,-m_1}^\dagger = \phi_{m_1,m_2}$.  This leads to the standard totally symmetric massive field having number of real-valued D.o.F equal to $(2j_1+1)^2$ .

\noindent \iiibf) Our study demonstrates that AdS space admits decoupled equations of motion for light-cone gauge fields. In this respect it will be interesting to find other gravitational backgrounds which admit decoupled equations of motion for massive fields. Discussion of massive fields in gravitational backgrounds may be found, e.g.,  in Refs.\cite{Buchbinder:1999ar,Cortese:2013lda}.

\noindent {\bf Mixed-symmetry massive fermionic fields}. Our results for bosonic fields are easily generalized to the case of fermionic fields. To this end, we introduce the following set of anticommuting  complex-valued fields
\be \label{12102014-43}
\psi_{m_1,m_2}\,, \qquad m_1=-j_1,-j_1+1,\ldots, j_1\,, \qquad m_2=-j_2,-j_2+1,\ldots, j_2\,.
\ee
By analogy with \rf{12102014-02}, fermionic fields in \rf{12102014-43} can be collected into a ket-vector defined by
\be \label{12102014-44}
\psik  = \sum_{m_1=-j_1}^{j_1}\sum_{m_2=-j_2}^{j_2} \frac{u_1^{j_1+m_1} v_1^{j_1-m_1} u_2^{j_2+m_2} v_2^{j_2-m_2}}{\sqrt{(j_1+m_1)!(j_1-m_1)!(j_2+m_2)!(j_2-m_2)!}} \psi_{m_1,m_2}|0\rangle\,.
\ee
Lagrangian for fermionic field is given by
\be \label{12102014-45}
\LL =  \psibr \frac{\irm }{\partial^+}\bigl(\Box + \partial_z^2 -\frac{1}{z^2}A\bigr) \psik\,,
\ee
where AdS mass operator $A$ entering Lagrangian \rf{12102014-45} takes the same form as the one for bosonic field in \rf{12102014-07}-\rf{12102014-11}.

\subsection{ \large Light-cone gauge realization of relativistic symmetries } \label{section-06}

Relativistic symmetries of fields in $AdS_5$ are described by the $so(4,2)$ algebra. This algebra contains the Lorentz sublagebra $so(3,1)$.
The choice of the light-cone gauge spoils the manifest $so(3,1)$ Lorentz symmetries. This is to say that, in the framework of the light-cone approach, complete description of field dynamics in $AdS_5$ implies that we have to work out realization of the $so(3,1)$ algebra symmetries and the remaining relativistic symmetries as well. We now turn to the discussion of the $so(4,2)$ algebra symmetries of the light-cone gauge action \rf{12102014-04}.

In light-cone approach, the $so(4,2)$ algebra  generators can be separated into two groups:
\beq
\label{12102014-15} && P^i,\ \ P^+, \ \ J^{+i}, \ \ J^{+-}, \ \ J^{ij}, \ \  D, \ \ K^i,\ \ K^+,\hspace{1cm} \hbox{ kinematical generators};\qquad
\\
\label{12102014-16} && P^-,\ \ J^{-i}\,, \ \ K^-\hspace{5.7cm} \hbox{ dynamical
generators},
\eeq
where vector indies of the $so(2)$ algebra take values $i,j=1,2$ (for more details, see Appendix A). Field theoretical representation of $so(4,2)$ algebra generators $G_\field$ in \rf{12102014-15}, \rf{12102014-16} takes the form
\be \label{12102014-17}
G_\field = \int dz dx^-d^2 x\,\langle\partial^+\phi|G_\diff|\phi\rangle + h.c.,
\ee
where $G_\diff$ stands for a realization of the generators in terms of differential operators acting on the bosonic ket-vector $\phik$ \rf{12102014-02}. We now present the explicit expressions for the differential operators $G_\diff$.

\noindent {\bf Kinematical generators},
\beq
\label{12102014-18} && P^i=\partial^i\,, \qquad P^+=\partial^+\,,
\\
\label{12102014-19} && J^{+-} = x^+P^-  - x^-\partial^+\,,
\\
\label{12102014-20} && J^{+i}=  x^+\partial^i - x^i\partial^+\,,
\\
\label{12102014-21} && J^{ij} = x^i\partial^j-x^j\partial^i + M^{ij}\,,
\\
\label{12102014-22} && D = x^+P^- + x^-\partial^+ + x^i\partial^i + z\partial_z + \frac{3}{2}\,,
\\
\label{12102014-23} && K^+ = -\frac{1}{2}(2x^+x^- + x^ix^i + z^2)\partial^+ +x^+D \,,
\\
\label{12102014-24} && K^i = -\frac{1}{2}(2x^+x^- + x^jx^j + z^2 )\partial^i + x^i D +
M^{ij} x^j + M^{i-}x^+ + M^{\ominussm i} \,,
\eeq

\noindent {\bf Dynamical generators},
\beq
\label{12102014-25} && P^-=\frac{-\partial^i\partial^i + \MM^2}{2\partial^+}\,,
\\
\label{12102014-26} && J^{-i}=x^-\partial^i-x^i P^- + M^{-i}\,,
\\
\label{12102014-27} && K^- = -\frac{1}{2}(2x^+x^-  + x^jx^j + z^2) P^- + x^-D
+ x^i M^{-i} - M^{\ominussm i} \frac{\partial^i}{\partial^+} +  \frac{1}{\partial^+} B \,, \qquad
\\
\label{12102014-27-a}&& \hspace{1cm}  \MM^2 \equiv -\partial_z^2 + \frac{1}{z^2} A\,,
\\
\label{12102014-28} && \hspace{1cm}  M^{-i} \equiv
M^{ij}\frac{\partial^j}{\partial^+}+\frac{1}{\partial^+}M^{\oplussm i}\,,
\\
\label{12102014-28-a}&&  \hspace{1cm} M^{\oplussm i}   =  - M^{zi}\partial_z - \frac{1}{2z} [M^{zi},A]\,,
\\
\label{12102014-29} && \hspace{1cm}  M^{\ominussm i}  =   - zM^{zi}\,.
\eeq
Spin operators $M^{ij}$, $M^{zi}$, $i,j=1,2$, form commutation relations of the $so(3)$  algebra
\beq
&& [M^{ij},M^{kl}]= \delta^{jk}M^{il} + 3 \hbox{ terms} \,, \qquad [M^{zi},M^{zj}] = - M^{ij}\,,
\\
&& [M^{ij}, M^{zk}]=\delta^{jk} M^{zi} - \delta^{ik} M^{zj}\,.
\eeq
Also, we note the following interesting commutator
\be
[M^{\oplussm i},M^{\oplussm j}] = \MM^2 M^{ij}\,.
\ee
Operator $A$ is presented in \rf{12102014-07}. All that remains to complete a description of generators in \rf{12102014-24}-\rf{12102014-27} is to find the operators $M^{zi}$, $M^{\oplussm i}$, $B$. In order to present these operators explicitly, we introduce a frame of complex coordinates $x^\Rsm$, $x^\Lsm$ defined by
\be
x^\Rsm \equiv \frac{1}{\sqrt{2}} (x^1 + \irm x^2)\,, \qquad
x^\Lsm \equiv \frac{1}{\sqrt{2}} (x^1 - \irm x^2)\,.
\ee
In such frame, a vector $X^i$ is decomposed as $X^i = X^\Rsm, X^\Lsm$, while a scalar product of two vectors $X^i$, $Y^i$ is represented as $X^i Y^i = X^\Rsm Y^\Lsm + X^\Lsm Y^\Rsm$.   This is to say that,  in a frame of the complex coordinates, the spin operator $M^{zi}$ is decomposed as $M^{zi} = M^{z\Rsm}, M^{z\Lsm}$,  while the  operator $M^{\oplussm i}$ is decomposed as $M^{\oplussm i} = M^{\oplussm \Rsm}, M^{\oplussm\Lsm}$. Note also that, in a frame of the complex coordinates, the $so(2)$ algebra generator $M^{ij}=-M^{ji}$ is represented as $M^{\Rsm\Lsm}$.  In a frame of the complex coordinates, we find the following representation for the operators entering generators of the $so(4,2)$ algebra in \rf{12102014-24}-\rf{12102014-27}:
\beq
\label{12102014-30} && M^{\Rsm\Lsm} = S_1 + S_2\,,
\\
\label{12102014-31} && M^{z\Rsm} = g_1 S_1^\Rsm + g_2 S_2^\Rsm\,,
\\
\label{12102014-32} && M^{z\Lsm} = - S_1^\Lsm g_1 - S_2^\Lsm g_2\,.
\\
\label{12102014-33} && M^{\oplussm \Rsm} = - \TT_{-\nu + \half} g_1 S_1^\Rsm - \TT_{\nu+\half} g_2 S_2^\Rsm\,,
\\
\label{12102014-34} && M^{\oplussm \Lsm} =    S_1^\Lsm g_1 \TT_{\nu - \half} +  S_2^\Lsm g_2  \TT_{-\nu-\half} \,,
\\
\label{12102014-34-a}&& B =  \kappa (S_1-S_2) + S_1^2 + S_2^2 - j_1(j_1+1) - j_2(j_2+1)\,,
\eeq
where we use the notation
\beq
\label{12102014-35} && \hspace{1cm} S_1^\Rsm  = \frac{1}{\sqrt{2}} u_1 \vb_1 \,,\qquad S_1^\Lsm  = \frac{1}{\sqrt{2}} v_1 \ub_1\,,
\\
\label{12102014-36} && \hspace{1cm} S_2^\Rsm  = \frac{1}{\sqrt{2}} u_2 \vb_2\,,\qquad S_2^\Lsm  = \frac{1}{\sqrt{2}} v_2 \ub_2\,,
\\
\label{12102014-37} && \hspace{1cm} g_1 = \Bigl(\frac{(\kappa + j_2
+S_1)(\kappa-j_2-1+S_1)}{(\kappa-1+S_1-S_2)(\kappa+S_1-S_2)} \Bigr)^{1/2}\,,
\\
\label{12102014-38} && \hspace{1cm} g_2 = \Bigl(\frac{(\kappa + j_1 + 1 -
S_2)(\kappa-j_1-S_2)}{(\kappa+1+S_1-S_2)(\kappa+S_1-S_2)}\Bigr)^{1/2}\,,
\\
\label{12102014-38-x1} && \hspace{1cm} \TT_\nu \equiv \partial_z + \frac{\nu}{z}\,.
\eeq
The following remarks are in order.

\noindent \ibf) Spin operators $S_\tau^{\Rsm,\Lsm}$ defined in \rf{12102014-35},\rf{12102014-36} and the ones in \rf{12102014-09} satisfy commutation relations of two $su(2)$ algebras
\be \label{12102014-39}
[S_\tau,S_\tau^\Rsm ] = S_\tau^\Rsm\,,\quad [S_\tau,S_\tau^\Lsm ] = - S_\tau^\Lsm\,, \quad [S_\tau^\Rsm , S_\tau^\Lsm] = S_\tau \,,  \quad
S_\tau^\dagger = S_\tau\,, \quad (S_\tau^\Rsm)^\dagger =  S_\tau^\Lsm ,\quad \tau =1,2\,.
\ee
In this paper, we use ket-vector \rf{12102014-02} which depends on the oscillators $u_\tau$, $v_\tau$. Realization of the operators $S_\tau$, $S_\tau^{\Rsm,\Lsm}$ on space of such ket-vector is given in \rf{12102014-09},\rf{12102014-35},\rf{12102014-36}. We note however that the above-given representation for the operators $A$, $B$, $M^{zi}$, $M^{ij}$ in terms of the operators $S_\tau$, $S_\tau^{\Rsm,\Lsm}$ does not depend on the particular form for the operators $S_\tau$, $S_\tau^{\Rsm,\Lsm}$  given in \rf{12102014-09},\rf{12102014-35},\rf{12102014-36}. What is important for our practical computations is that the operators $S_\tau$, $S_\tau^{\Rsm,\Lsm}$ satisfy the relations given in \rf{12102014-39}.

\noindent \iibf) The light-cone gauge action \rf{12102014-04} is invariant under the following $so(4,2)$ algebra transformations:
\be \label{12102014-40}
\delta_G \phik = G_\diff \phik\,.
\ee

\noindent \iiibf) Lagrangian \rf{12102014-05} implies the standard equal-time Poisson-Dirac bracket,
\be \label{12102014-41}
\bigl[\, |\phi(x,z)\rangle,\langle\phi(x',z')| \, \bigr]\Bigr|_{{\rm equal} \ x^+} = - \frac{1}{2\partial^+}
\delta^{(2)}(x-x')\delta(x^--x'{}^-)\delta(z-z') |\rangle \langle |  \,,
\ee
where a dependence of the ket-vectors on the coordinates of $AdS_5$ space is shown explicitly and the notation $|\rangle\langle|$ stands for the corresponding unit operator on space of  ket-vector \rf{12102014-02}.

\noindent \ivbf) Using \rf{12102014-41}, we verify that equal time commutator of the ket-vector $\phik$ with the field theoretical generator \rf{12102014-17} takes the form
\be  \label{12102014-42}
[|\phi\rangle,G_\field] = G_\diff|\phi\rangle,
\ee
as it should be.

\noindent \vbf)  Defining an action of the operators $M^{zi}$ on the component fields by the relation
\be
M^{zi} \phik  = \sum_{m_1=-j_1}^{j_1}\sum_{m_2=-j_2}^{j_2} \frac{u_1^{j_1+m_1} v_1^{j_1-m_1} u_2^{j_2+m_2} v_2^{j_2-m_2}}{\sqrt{(j_1+m_1)!(j_1-m_1)!(j_2+m_2)!(j_2-m_2)!}} (M^{zi} \phi_{m_1,m_2})|0\rangle\,,
\ee
we get the following transformation rules of the component fields $\phi_{m_1,m_2}$ in \rf{12102014-01},
\beq
\label{13102014-01d} M^{z\Rsm}\phi_{_{m_1,m_2}} & = & \Bigl(\frac{(\kappa+j_2+m_1)(\kappa-1-j_2+m_1)}{
(\kappa-1+m_1-m_2)(\kappa+m_1-m_2)}\Bigr)^{1/2} r_{j_1,m_1} \phi_{_{m_1-1,m_2}}
\nonumber\\
& +  & \Bigl(\frac{(\kappa+1+j_1-m_2)(\kappa-j_1-m_2)}{ (\kappa+1+
m_1-m_2)(\kappa+m_1-m_2)}\Bigr)^{1/2} r_{j_2,m_2} \phi_{_{m_1,m_2-1}}\,,
\\[15pt]
\label{13102014-02d} -M^{z\Lsm}\phi_{_{m_1,m_2}} & = & \Bigl(\frac{(\kappa+1+j_2+m_1)(\kappa-j_2+m_1)}{
(\kappa+m_1-m_2)(\kappa+1+m_1-m_2)}\Bigr)^{1/2} r_{j_1,m_1+1} \phi_{_{m_1+1,m_2}}
\nonumber\\
& +  & \Bigl(\frac{(\kappa+ j_1-m_2)(\kappa-1-j_1-m_2)}{ (\kappa+
m_1-m_2)(\kappa-1+m_1-m_2)}\Bigr)^{1/2} r_{j_2,m_2+1} \phi_{_{m_1,m_2+1}}\,,
\\
&& r_{j,m} \equiv \sqrt{(j+m)(j-m+1)/2}\,.
\eeq

\noindent \vibf) Expressions for operators $A$, $B$, $M^{zi}$ above given are obtained by solving the defining equations (A1)-(A4) given in Appendix A in Ref.\cite{Metsaev:2003cu}.

\noindent {\bf Relativistic symmetries of fermionic fields}. Lagrangian for fermionic field \rf{12102014-45} is invariant under transformations of the $so(4,2)$ algebra given by
\be  \label{13102014-01}
\delta_G\psik = G_\diff^\ferm \psik\,,
\ee
where differential operators of fermionic fields $G_\diff^\ferm$ can be obtained from those
of bosonic fields \rf{12102014-18}-\rf{12102014-27} by making there the following
substitution
\be  \label{13102014-01a}
x^- \rightarrow x^- +\frac{1}{2\partial^+}\,.
\ee
Lagrangian \rf{12102014-45} implies the standard equal-time Poisson-Dirac antibracket,
\be \label{13102014-02}
\bigl\{\, |\psi(x,z)\rangle,\langle\psi(x',z')| \,\bigr\}\Bigr|_{{\rm equal} \ x^+} = \frac{\irm }{2}
\delta^{(2)}(x-x')\delta(x^--x'{}^-) \delta(z-z') |\rangle \langle |  \,,
\ee
where  the notation $|\rangle\langle|$ stands for the corresponding unit operator on space of  ket-vector \rf{12102014-44}.

For the fermionic field, field theoretical representation of the $so(4,2)$ algebra generators $G_\field$ in \rf{12102014-15}, \rf{12102014-16} takes the form
\be  \label{13102014-03}
G_\field^\ferm = -\irm  \int dz dx^- d^2 x  \psibr G_\diff^\ferm \psik + h.c.\,,
\ee
Using \rf{13102014-02},\rf{13102014-03}, we verify that equal-time commutator of the ket-vector $\psik$ with field theoretical generator \rf{13102014-03} takes the form
\be
[\psik,G_\field^\ferm] = G_\diff^\ferm\psik.
\ee

\newsection{Massless mixed-symmetry fields in $AdS_5$}\label{masless}

In Sec.\ref{mix-lagr}, we found the light-cone gauge action for the mixed-symmetry massive fields. For the massive field, the lowest eigenvalue of the energy operator should satisfy the restriction $E_0 > h_1 + 2$ (see \rf{13102014-17}). To realize the limit of the mixed-symmetry massless field, we take
\be \label{13102014-04}
E_0 \rightarrow h_1 +2\,,\qquad  h_1 > |h_2|\,.
\ee
In terms of labels $\kappa$, $j_1$, $j_2$, \rf{13102014-18-x1}, limit in \rf{13102014-04} can equivalently be represented as
\be \label{13102014-05}
\kappa \rightarrow j_1 + j_2\,,\qquad  j_1 \ne 0 \,, \quad j_2 \ne 0\,.
\ee
Below, we will demonstrate, that limit \rf{13102014-04} leads to appearance of invariant subspace in $\phik$ \rf{12102014-02}. Namely we state that, in the limit \rf{13102014-04}, the following set of fields
\beq
&& \phi_{-j_1,j_2}\,,
\nonumber\\
&& \phi_{m,j_2}\,, \qquad   \ \  m = -j_1+1,-j_1+2,\ldots , j_1\,,
\nonumber\\
\label{13102014-06} && \phi_{-j_1,m}\,, \qquad m=-j_2,-j_2+1,\ldots, j_2-1\,,
\eeq
are invariant under action of $so(4,2)$ algebra generators given in \rf{12102014-18}-\rf{12102014-27}. In other words, in limit \rf{13102014-04}, fields \rf{13102014-06} transform into themselves under the action of the generators given in \rf{12102014-18}-\rf{12102014-27}. Before proving this statement let us present our result for Lagrangian of mixed-symmetry massless field.

\noindent {\bf \bf Lagrangian for mixed-symmetry massless field}. Equating to zero all component fields $\phi_{m_1,m_2}$ in \rf{12102014-02} with the exception of ones appearing in \rf{13102014-06}, we note that ket-vector which describes massless mixed-symmetry field is given by
\beq
\label{13102014-07} \phik & = & |\phi_0\rangle + |\phi_1\rangle + |\phi_2\rangle\,,
\\
\label{13102014-07-a1} |\phi_0\rangle  &  = & \frac{v_1^{2j_1}  u_2^{2j_2} }{\sqrt{(2j_1)!(2j_2)!}}\, \phi_{-j_1,j_2}|0\rangle\,,
\nonumber\\
|\phi_1\rangle &= & \sum_{m=-j_1+1}^{j_1} \frac{u_1^{j_1+m} v_1^{j_1-m} u_2^{2j_2} }{\sqrt{(j_1+m)!(j_1-m)!(2j_2)!}}\, \phi_{m,j_2}|0\rangle\,,
\nonumber\\
|\phi_2\rangle & = &  \sum_{m=-j_2}^{j_2-1} \frac{  v_1^{2j_1} u_2^{j_2+m} v_2^{j_2-m} }{\sqrt{(2j_1)!(j_2+m)!(j_2-m)!}}\, \phi_{-j_1,m}|0\rangle\,.
\eeq
Plugging \rf{13102014-07} into Lagrangian \rf{12102014-05} and using $\kappa$ given in \rf{13102014-05}, we find the following Lagrangian:
\beq
\label{13102014-08-c1} && \hspace{-2.3cm} \LL  =  \LL_0 + \LL_1 + \LL_2\,,
\\
\LL_0  &  = & \langle \phi_0| (\Box + \partial_z^2 + \frac{1}{4z^2}) |\phi_0\rangle\,,
\nonumber\\
\LL_1 &= &  \langle\phi_1| \Bigl( \Box + \partial_z^2 - \frac{1}{z^2}\bigl((j_1+S_1)^2 -\frac{1}{4})\Bigr)|\phi_1\rangle\,,
\nonumber\\
\label{13102014-08} \LL_2 &= &  \langle\phi_2| \Bigl( \Box + \partial_z^2 - \frac{1}{z^2}\bigl((j_2 - S_2)^2 -\frac{1}{4})\Bigr)|\phi_2\rangle\,,
\eeq
where $S_1$, $S_2$ are given in \rf{12102014-09}. In terms of component fields \rf{13102014-06}, expressions for $\LL_0$, $\LL_1$, $\LL_2$ \rf{13102014-08} can be represented as
\beq
\LL_0  &  = & \phi_{-j_1,j_2}^\dagger (\Box + \partial_z^2 + \frac{1}{4z^2})\phi_{-j_1,j_2}\,,
\nonumber\\
\LL_1 &= & \sum_{m=-j_1+1}^{j_1} \phi_{m,j_2}^\dagger \Bigl( \Box + \partial_z^2 - \frac{1}{z^2}\bigl((j_1+m)^2 -\frac{1}{4})\Bigr) \phi_{m,j_2}\,,
\nonumber\\
\label{13102014-08-c2} \LL_2 &= & \sum_{m=-j_2+1}^{j_2} \phi_{-j_1,-m}^\dagger \Bigl( \Box + \partial_z^2 - \frac{1}{z^2}\bigl((j_2+m)^2 -\frac{1}{4})\Bigr) \phi_{-j_1,-m}\,.
\eeq

\noindent {\bf Transformations of mixed-symmetry massless fields}. Now we demonstrate that, in limit \rf{13102014-04}, fields \rf{13102014-06} transform into themselves under the action of the generators given in \rf{12102014-18}-\rf{12102014-27}. From \rf{12102014-18}-\rf{12102014-27}, we see that we should verify that fields \rf{13102014-06} transform into themselves under the action of the operators $A$, $B$, $M^{\Rsm\Lsm}$, $M^{zi}$, $M^{\oplussm i}$. Using expressions for operators $A$ \rf{12102014-07}, $B$ \rf{12102014-34-a}, and $M^{\Rsm\Lsm}$ \rf{12102014-30}, we see that these operators are diagonal on space of the fields $\phi_{m_1,m_2}$. Taking this into account and using \rf{12102014-28-a}, we see that if fields \rf{13102014-06} are invariant under the action of the operator $M^{zi}$ then they also are invariant under the action of the operator $M^{\oplussm i}$.

Thus all that remains is to check that, in limit \rf{13102014-04}, fields \rf{13102014-06} transform into themselves under the action of the operator $M^{zi}$.
To this end we use \rf{13102014-01d}, \rf{13102014-02d} and note that, under the action of the operators $M^{z\Rsm}$, $M^{z\Lsm}$, fields \rf{13102014-06} transform as
\beq
\label{11102014-01} M^{z\Rsm}\phi_{_{m_1,j_2}} & = & \Bigl(\frac{\kappa+j_2+m_1}{
\kappa-j_2+m_1}\Bigr)^{1/2} r_{j_1,m_1} \phi_{_{m_1-1,j_2}}
\nonumber\\
& +  & \Bigl(\frac{(\kappa+1+j_1-j_2)(\kappa-j_1-j_2)}{ (\kappa+1-j_2+
m_1)(\kappa-j_2+m_1)}\Bigr)^{1/2} r_{j_2,j_2} \phi_{_{m_1,j_2-1}}\,,
\nonumber\\
&& m_1 = -j_1+1,-j_1+2,\ldots, j_1\,,
\\
M^{z\Rsm}\phi_{_{-j_1,m_2}} & = & \Bigl(\frac{\kappa+1+j_1-m_2}{ \kappa + 1 -j_1-m_2}\Bigr)^{1/2} r_{j_2,m_2} \phi_{_{-j_1,m_2-1}}\,,
\nonumber\\
&& m_2 = -j_2,-j_2+1,\ldots, j_2\,,
\\[10pt]
\label{11102014-03} -M^{z\Lsm}\phi_{_{-j_1,m_2}} & = & \Bigl(\frac{(\kappa+1+j_2-j_1)(\kappa-j_1-j_2)}{
(\kappa - j_1 - m_2)(\kappa+1 - j_1-m_2)}\Bigr)^{1/2} r_{j_1,-j_1+1} \phi_{_{-j_1+1,m_2}}
\nonumber\\
& +  & \Bigl(\frac{\kappa+ j_1-m_2}{ \kappa -
j_1-m_2}\Bigr)^{1/2} r_{j_2,m_2+1} \phi_{_{-j_1,m_2+1}}\,,
\nonumber\\
&& m_2 = -j_2,-j_2+1,\ldots, j_2-1\,,
\\[10pt]
\label{11102014-04} -M^{z\Lsm}\phi_{_{m_1,j_2}} & = & \Bigl(\frac{\kappa+1+j_2+m_1}{\kappa+1-j_2+m_1}\Bigr)^{1/2} r_{j_1,m_1+1} \phi_{_{m_1+1,j_2}}\,,
\nonumber\\
&& m_1 = -j_1,-j_1+1,\ldots, j_1\,.
\eeq
From \rf{11102014-01}, \rf{11102014-03}, we see that, for arbitrary $\kappa$, fields in  \rf{13102014-06} are not invariant under the action of the operators $M^{z\Rsm}$, $M^{z\Lsm}$. However, we note that, in the massless limit,
\be
\kappa \rightarrow j_1 + j_2\,,
\ee
relations \rf{11102014-01}-\rf{11102014-04} take the form
\beq
\label{11102014-05} M^{z\Rsm}\phi_{_{m_1,j_2}} & = & \Bigl(\frac{j_1+2j_2+m_1}{
j_1+m_1}\Bigr)^{1/2} r_{j_1,m_1} \phi_{_{m_1-1,j_2}}\,,
\nonumber\\
&& m_1 = -j_1+1,-j_1+2,\ldots, j_1\,,
\\
M^{z\Rsm}\phi_{_{-j_1,m_2}} & = & \Bigl(\frac{2j_1 +j_2+1-m_2}{ j_2 + 1 - m_2}\Bigr)^{1/2} r_{j_2,m_2} \phi_{_{-j_1,m_2-1}}\,,
\nonumber\\
&& m_2 = -j_2,-j_2+1,\ldots, j_2\,,
\\[10pt]
-M^{z\Lsm}\phi_{_{-j_1,m_2}} & =&  \Bigl(\frac{2j_1 + j_2 -m_2}{
j_2-m_2}\Bigr)^{1/2} r_{j_2,m_2+1} \phi_{_{-j_1,m_2+1}}\,,
\nonumber\\
&& m_2 = -j_2,-j_2+1,\ldots, j_2-1\,,
\\[10pt]
\label{11102014-08} -M^{z\Lsm}\phi_{_{m_1,j_2}} & = & \Bigl(\frac{j_1 + 2j_2+1 + m_1}{j_1+1+m_1}\Bigr)^{1/2} r_{j_1,m_1+1} \phi_{_{m_1+1,j_2}}
\nonumber\\
&& m_1 = -j_1,-j_1+1,\ldots, j_1\,.
\eeq
From \rf{11102014-05}-\rf{11102014-08}, we see that, in massless limit, fields
in \rf{13102014-06} are indeed invariant under the action of the operators $M^{z\Rsm}$, $M^{z\Lsm}$.

\noindent {\bf Lagrangian for totally symmetric massless fields}. For totally symmetric bosonic fields, we have the condition $j_1 = j_2$. Using the notation $j\equiv j_1$, we can impose the reality condition
\beq
\label{20102014-10} && \phi_{-j,j}^\dagger = \phi_{-j,j}\,,
\\
 \label{20102014-11} && \phi_{m,j}^\dagger = \phi_{-j,-m}\,, \qquad m= -j+1, -j+2, \ldots , j\,.
\eeq
Introducing then the real-valued field $\phi_0$, $\phi_0^\dagger=\phi_0$, and a set of the complex-valued fields $\phi_m$ by the relations
\be  \label{20102014-01}
\phi_0 \equiv \sqrt{2} \phi_{-j,j}, \qquad \phi_{j+m} \equiv \sqrt{2} \phi_{m,j}\,, \qquad m =-j+1,-j+2,\ldots j\,,
\ee
we see that Lagrangian \rf{13102014-08-c1},\rf{13102014-08-c2} takes the form
\be \label{20102014-02}
\LL =  \half \phi_0 \bigl( \Box + \partial_z^2 + \frac{1}{4z^2} \bigr) \phi_0
+ \sum_{m=1}^{2j} \phi_m^\dagger \Bigl( \Box + \partial_z^2 - \frac{1}{z^2}\bigl(m^2 -\frac{1}{4})\Bigr) \phi_m\,.
\ee
Number of propagating real-valued D.o.F. entering Lagrangian \rf{20102014-02} is equal to $2s+1$, where $s\equiv 2j$. Lagrangian \rf{20102014-02} describes light-cone gauge dynamics of spin-$s$ totally symmetric massless field in $AdS_5$.

\noindent {\bf Fermionic mixed-symmetry and totally symmetric massless fields}.  To get Lagrangian for mixed-symmetry massless fermionic field we use a ket-vector $\psik$ which is obtained from \rf{13102014-07} by substituting anticommuting fields $\psi_{-j_1,j_2}$,  $\psi_{m,j_2}$, $\psi_{-j_1,m}$ in the respective ket-vectors in \rf{13102014-07-a1}. Plugging then such ket-vector $\psik$ into \rf{12102014-45} and setting $\kappa=j_1+j_2$, we get Lagrangian for mixed-symmetry massless fermionic field.
Setting $j_1=\half (h_1+\half)$, $j_2=\half (h_1-\half)$ in such Lagrangian for mixed-symmetry fermionic field, we get Lagrangian for totally symmetric massless fermionic field.

\newsection{Self-dual massive fields in $AdS_5$}\label{self-dual}

The case of self-dual massive fields \rf{13102014-05} is realized by
considering
\be \label{13102014-20}
E_0 > h_1+1\,, \qquad h_1  = |h_2|>1/2\,.
\ee
Using labels $\kappa$, $j_1$, $j_2$ \rf{13102014-18-x1} and considering $h_2>1/2$, we represent \rf{13102014-20} as
\be \label{13102014-21}
\kappa > j_1 - 1\,, \qquad j_2  = 0\,, \quad \hbox{ for } \quad h_2 > 1/2\,.
\ee
Using the relation $j_2=0$ in \rf{12102014-02}, we see that ket-vector of the bosonic field \rf{12102014-02} simplifies as
\be \label{13102014-22}
\phik  = \sum_{m=-j_1}^{j_1} \frac{u_1^{j_1 + m} v_1^{j_1 - m}}{\sqrt{(j_1+m)!(j_1-m)!}} \phi_{m,0}|0\rangle\,.
\ee
Ket-vector \rf{13102014-22} describes the massive self-dual field. Plugging this ket-vector  into \rf{12102014-05}, we obtain Lagrangian of the massive self-dual field with the operator $A$ given by
\be \label{13102014-23}
A = \nu^2 - \frac{1}{4}\,, \qquad \nu = \kappa + S_1 \,, \qquad \kappa \equiv E_0-2\,,
\ee
where $S_1$ is given in \rf{12102014-09}. In terms of the component fields $\phi_{m,0}$,  Lagrangian of the massive self-dual field takes the form
\be
\LL=  \sum_{m = -j_1}^{j_1} \phi_{m,0}^\dagger \Bigl( \Box + \partial_z^2 - \frac{1}{z^2}\bigl((\kappa + m)^2 -\frac{1}{4})\Bigr) \phi_{m,0}\,.
\ee
Note that the relation $j_2=0$ implies the relations $S_2\phik=0$, $S_2^{\Rsm,\Lsm}\phik=0$. Using these relations in \rf{12102014-30}-\rf{12102014-34-a}, we get operators entering generators of the $so(4,2)$ algebra  \rf{12102014-18}-\rf{12102014-27} corresponding to the self-dual massive field,
\beq
\label{13102014-24} && M^{\Rsm\Lsm} = S_1\,, \qquad  M^{z\Rsm} = S_1^\Rsm\,, \qquad M^{z\Lsm} = - S_1^\Lsm\,,
\\
\label{13102014-25} && M^{\oplussm \Rsm} = - \TT_{-\nu + \half} S_1^\Rsm \,,  \qquad M^{\oplussm \Lsm} =    S_1^\Lsm  \TT_{\nu - \half}\,,\qquad \TT_\nu \equiv \partial_z + \frac{\nu}{z}\,,
\\
\label{13102014-26} && B =  \kappa S_1 + S_1^2 - j_1(j_1+1)\,.
\eeq

\noindent {\bf Fermionic self-dual massive fields}. Our result for bosonic fields above presented can be extended to the case of fermionic fields is a rather straightforward way. For this case, we use the fermionic ket-vector given by
\be \label{14102014-01}
\psik  = \sum_{m=-j_1}^{j_1} \frac{u_1^{j_1 + m} v_1^{j_1 - m}}{\sqrt{(j_1+m)!(j_1-m)!}} \psi_{m,0}|0\rangle\,,
\ee
while the operator $A$ takes the form given in \rf{13102014-23}.  Plugging \rf{14102014-01} and $A$ \rf{13102014-23} into \rf{12102014-45}, we get Lagrangian of fermionic self-dual massive field
\be \label{14102014-02}
\LL=  \sum_{m = -j_1}^{j_1} \psi_{m,0}^\dagger\, \frac{\irm }{\partial^+}\Bigl( \Box + \partial_z^2 - \frac{1}{z^2}\bigl((\kappa + m)^2 -\frac{1}{4})\Bigr) \psi_{m,0}\,.
\ee
Operators for bosonic fields in \rf{13102014-24}-\rf{13102014-26} take the same form for fermionic self-dual massive fields.

\newsection{ Mixed-symmetry anomalous currents and shadows in $R^{3,1}$} \label{mix-cur-sh}

Let us recall some basic notions of CFT. Fields of CFT can be separated into two groups: currents and shadows. Currents and shadows in $R^{3,1}$ transform in representations of conformal algebra $so(4,2)$. Currents transform in representations of $so(4,2)$ labelled by $\Delta_\cur$, $h_1$, $h_2$, where $\Delta_\cur$ is conformal dimension, while $h_1$ and $h_2$ are highest weights for representations of the $so(4)$ algebra.
In this paper, currents having conformal dimensions $\Delta_\cur = E_0$ with $E_0$ as in \rf{13102014-15}, \rf{13102014-16}, and \rf{13102014-17} are referred to as self-dual current, mixed-symmetry canonical current, and mixed-symmetry anomalous current respectively. Shadows transform in representations of $so(4,2)$ labelled by $\Delta_\sh$, $h_1$, $h_2$, where $\Delta_\sh$ is conformal dimension, while $h_1$ and $h_2$ are highst weights for representations of the $so(4)$ algebra. In this paper, shadows having conformal dimensions $\Delta_\sh = 4-E_0$ with $E_0$ as in \rf{13102014-15}, \rf{13102014-16}, and \rf{13102014-17} are referred to as self-dual shadow, mixed-symmetry canonical shadow, and mixed-symmetry anomalous shadow respectively. Often, in place of labels $E_0$, $h_1$, $h_2$, we will use the labels $\kappa$, $j_1$, $j_2$ defined in \rf{13102014-18-x1}. We start our light-cone gauge formulation of currents and shadows with the description of field contents.%
\footnote{ In the framework of light-cone approach, the fields we use for the description of currents and shadows are not subject to any differential constraints. We recall that, in the framework of Lorentz covariant approaches, fields which are used for description of currents are subject to differential constraints. Study of differential constraints for mixed-symmetry currents may be found in Refs.\cite{Shaynkman:2004vu,Dobrev:2012mea,Alkalaev:2012rg}.}

\noindent {\bf Field content of mixed-symmetry anomalous current and anomalous shadow}. To discuss light-cone gauge formulation of bosonic mixed-symmetry anomalous current and
anomalous shadow we use the following complex-valued fields in $R^{3,1}$:
\be \label{17102014-01}
\phi_{\cur;m_1,m_2}\,,\qquad \phi_{\sh;m_1,m_2}\,, \qquad m_1=-j_1,-j_1+1,\ldots, j_1\,, \qquad m_2=-j_2,-j_2+1,\ldots, j_2\,. %
\ee
In order to streamline the presentation of light-cone gauge description,
we use oscillators $u_\tau$, $v_\tau$ \rf{26102014-01},\rf{26102014-02} and collect fields \rf{17102014-01} into ket-vector $|\phi_\cur\rangle$,  $|\phi_\sh\rangle$ defined by the relation
\be \label{17102014-02}
|\phi_{\cur,\sh}\rangle  = \sum_{m_1=-j_1}^{j_1}\sum_{m_2=-j_2}^{j_2} \frac{u_1^{j_1+m_1} v_1^{j_1-m_1} u_2^{j_2+m_2} v_2^{j_2-m_2}}{\sqrt{(j_1+m_1)!(j_1-m_1)!(j_2+m_2)!(j_2-m_2)!}} \phi_{\cur,\sh;m_1,m_2}|0\rangle\,.
\ee
Conformal dimensions of the fields in \rf{17102014-01} are given by
\be \label{17102014-03}
\Delta(\phi_{\cur;m_1,m_2})=  2 + \kappa +
m_1 - m_2\,,\qquad \Delta(\phi_{\sh;m_1,m_2})= 2 - \kappa
- m_1 + m_2\,.
\ee

\noindent {\bf Light-cone gauge 2-point vertices of mixed-symmetry anomalous currents and shadows}. For currents and shadows, one can construct two 2-point vertices. The first 2-point vertex, which we denote by $\Gamma^{\rm cur-\sh}$, is a local functional of current and shadow. Using notation $|\phi_\cur\rangle$ and $|\phi_\sh\rangle$ for the respective ket-vectors of currents and shadows given in \rf{17102014-02}, we note the following expression for the local vertex:
\be \label{16102014-18}
\Gamma^{\rm cur-\sh} = \int d^4 x \, \LL^{\rm cur-sh}\,, \qquad \LL^{\rm cur-sh} = \langle \phi_\cur|| \phi_\sh\rangle \,.
\ee

The second 2-point vertex, which we denote by $\Gamma^{\rm sh-sh}$, is a non-local functional of shadows. Using notation $|\phi_\sh\rangle$ for the ket-vector of shadow \rf{17102014-02}, we note the following expression for the non-local vertex:
\beq
\label{16102014-19} && \Gamma^{\rm sh-sh}= \int d^4 x_1 d^4 x_2 \, \LL^{\rm sh-sh} \,,
\\
\label{16102014-19-a1}&& \hspace{1.3cm} \LL^{\rm sh-sh} \equiv  \langle\phi_\sh(x_1)|
\frac{f_\nu}{ |x_{12}|^{2\nu + 4 }} |\phi_\sh (x_2)\rangle \,,
\\
&& \hspace{1.3cm} f_\nu \equiv \frac{4^\nu \Gamma(\nu + 2)\Gamma(\nu + 1)}{4^\kappa
\Gamma(\kappa + 2)\Gamma(\kappa + 1)} \,,
\\
&& \hspace{1.3cm} |x_{12}|^2 \equiv x_{12}^a x_{12}^a\,, \qquad x_{12}^a = x_1^a - x_2^a\,.
\eeq
In terms of component fields \rf{17102014-01}, light-cone gauge 2-point vertex  \rf{16102014-19} can be represented as
\beq
\label{20102014-03} && \LL^{\rm sh-sh}  =   \sum_{m_1=-j_1}^{j_1} \sum_{m_2=-j_2}^{j_2} \phi_{\sh;\,m_1,m_2}^\dagger \frac{f_{\nu_{m_1,m_2} }}{ |x_{12}|^{2\nu_{m_1,m_2} + 4 }} \phi_{\sh;\,m_1,m_2}\,.
\\
\label{20102014-04} &&  f_{\nu_{m_1,m_2}} \equiv \frac{4^{\nu_{m_1,m_2}} \Gamma(\nu_{m_1,m_2} + 2)\Gamma(\nu_{m_1,m_2} + 1)}{4^\kappa
\Gamma(\kappa + 2)\Gamma(\kappa + 1)}\,,\qquad \nu_{m_1,m_2}\equiv \kappa + m_1 - m_2\,.\qquad
\eeq

\noindent {\bf Light-cone gauge 2-point vertex for mixed-symmetry canonical shadow}. All that is required to get 2-point vertex $\LL^{\sh-\sh}$ for canonical shadow is to set $\kappa = j_1 + j_2$ and plug ket-vector corresponding to canonical shadow into \rf{16102014-19-a1}. For canonical shadow, an expansion of ket-vector into component fields takes the same form as for massless AdS field in \rf{13102014-07}. This is to say that ket-vector of the mixed-symmetry canonical shadow can be presented in terms of the component fields as
\beq
\label{20102014-05} && \hspace{-2cm} |\phi_\sh\rangle  =  |\phi_{\sh,\,0}\rangle + |\phi_{\sh,\,1}\rangle + |\phi_{\sh,\,2}\rangle\,,
\\
|\phi_{\sh,\, 0}\rangle  &  = & \frac{v_1^{2j_1}  u_2^{2j_2} }{\sqrt{(2j_1)!(2j_2)!}}\, \phi_{\sh;\,-j_1,j_2}|0\rangle\,,
\nonumber\\
|\phi_{\sh,\,1}\rangle &= & \sum_{m=-j_1+1}^{j_1} \frac{u_1^{j_1+m} v_1^{j_1-m} u_2^{2j_2} }{\sqrt{(j_1+m)!(j_1-m)!(2j_2)!}}\, \phi_{\sh;\,m,j_2}|0\rangle\,,
\nonumber\\
|\phi_{\sh,\,2}\rangle & = &  \sum_{m=-j_2}^{j_2-1} \frac{  v_1^{2j_1} u_2^{j_2+m} v_2^{j_2-m} }{\sqrt{(2j_1)!(j_2+m)!(j_2-m)!}}\, \phi_{\sh;\,-j_1,m}|0\rangle\,.
\eeq
Plugging \rf{20102014-05} into \rf{16102014-19-a1} and using $\kappa$ given in \rf{13102014-05}, we find the following 2-point vertex:
\beq
\label{20102014-06} && \LL^{\sh-\sh} = \LL_0^{\sh-\sh} + \LL_1^{\sh-\sh} + \LL_2^{\sh-\sh}\,,
\\
\LL_0^{\sh-\sh}  &  = & \phi_{\sh;\,-j_1,j_2}^\dagger  \frac{ f_0 }{ |x_{12}|^4 } \phi_{\sh;\,-j_1,j_2}\,,
\nonumber\\
\LL_1^{\sh-\sh} &= & \sum_{m=-j_1+1}^{j_1} \phi_{\sh;\,m,j_2}^\dagger \frac{ f_{j_1 + m} }{ |x_{12}|^{ 2j_1 + 2m +4} }   \phi_{\sh;\,m,j_2}\,,
\nonumber\\
\LL_2^{\sh-\sh} &= & \sum_{m=-j_2+1}^{j_2} \phi_{\sh;\,-j_1,-m}^\dagger \frac{ f_{j_2 + m} }{ |x_{12}|^{ 2j_2 + 2m +4} }  \phi_{\sh;\,-j_1,-m}\,,
\\
&&  f_m \equiv \frac{ 4^m \Gamma(m + 2)\Gamma(m + 1)}{4^{j_1+j_2}
\Gamma(j_1+j_2 + 2)\Gamma(j_1+j_2 + 1)} \,.\qquad
\eeq
Expression for $ \LL^{\sh-\sh}$ in \rf{20102014-06} provides us light-cone gauge 2-point vertex for the canonical mixed-symmetry shadow in $R^{3,1}$ with arbitrary labels $j_1$, $j_2$.

\noindent {\bf Light-cone gauge 2-point vertex of self-dual shadow}. In order to get 2-point vertex $\LL^{\sh-\sh}$ for the self-dual shadow, we set $j_2=0$ and plug ket-vector corresponding to the self-dual shadow into \rf{16102014-19-a1}. For the self-dual  shadow, expansion of ket-vector into component fields takes the same form as for the massive self-dual AdS field in \rf{13102014-22}. This is to say that ket-vector of the self-dual shadow can be presented in terms of the component fields as
\be \label{20102014-07}
|\phi_\sh\rangle   = \sum_{m=-j_1}^{j_1} \frac{u_1^{j_1 + m} v_1^{j_1 - m}}{\sqrt{(j_1+m)!(j_1-m)!}} \phi_{\sh,\,m,0}|0\rangle\,.
\ee
Plugging \rf{20102014-07} into \rf{16102014-19-a1} and using the self-duality restrictions $j_2=0$, $S_2=0$, we find the following 2-point vertex:
\beq
\label{20102014-08} && \LL^{\sh-\sh} =  \sum_{m = -j_1}^{j_1} \phi_{m,0}^\dagger \frac{ f_{\nu_{m,0}} }{ |x_{12}|^{ 2\nu_{m,0} + 4} } \phi_{m,0}\,.
\\
&&  f_{\nu_{m,0}} \equiv \frac{4^{\nu_{m,0}} \Gamma(\nu_{m,0} + 2)\Gamma(\nu_{m,0} + 1)}{4^\kappa
\Gamma(\kappa + 2)\Gamma(\kappa + 1)}\,,\qquad \nu_{m,0}\equiv \kappa + m\,.\qquad
\eeq
Expression for $ \LL^{\sh-\sh}$ in \rf{20102014-08} provides us light-cone gauge 2-point vertex of the self-dual shadow in $R^{3,1}$ with arbitrary labels $\kappa$ and $j_1$.

\noindent {\bf Light-cone gauge 2-point vertex for totally symmetric canonical shadow}. For totally symmetric bosonic shadow, we have the condition $j_1 = j_2$. Using the notation $j\equiv j_1$, we can impose the reality condition on the shadows as in \rf{20102014-10}, \rf{20102014-11}. We introduce then real-valued field $\phi_{\sh,\,0}$, $\phi_{\sh,\,0}^\dagger=\phi_{\sh,\,0}$, and a set of complex-valued fields $\phi_{\sh,\,m}$ by the relations
\be \label{20102014-14}
\phi_{\sh,\,0} \equiv \sqrt{2} \phi_{\sh;\,-j,j}, \qquad \phi_{\sh,\,j+m} \equiv \sqrt{2} \phi_{\sh;\,m,j}\,, \qquad m =-j+1,-j+2,\ldots j\,.
\ee
Using \rf{20102014-14} in \rf{20102014-06} and setting $j_1=j$, $j_2=j$, we see that vertex \rf{20102014-06} takes the form
\beq
\label{20102014-14-a1} && \LL^{\sh-\sh} = \half \phi_{\sh, 0}  \frac{ f_0 }{ |x_{12}|^4 } \phi_{\sh, \, 0}
+ \sum_{m=1}^{2j} \phi_{\sh,\,m}^\dagger \frac{ f_m }{ |x_{12}|^{2m +4} }   \phi_{\sh,\,m}\,,
\\
&&  f_m \equiv \frac{ 4^m \Gamma(m + 2)\Gamma(m + 1)}{4^{2j}
\Gamma(2j + 2)\Gamma(2j + 1)} \,,\qquad m=0,1,\ldots, 2j.
\eeq
We note that light-cone gauge vertices for totally symmetric arbitrary spin canonical and anomalous shadows in $R^{d-1,1}$, with arbitrary $d$ were obtained in Ref.\cite{Metsaev:2009ym} and Ref.\cite{Metsaev:2011uy} respectively. In Refs.\cite{Metsaev:2009ym,Metsaev:2011uy}, gauge invariant and manifestly Lorentz invariant representation of vertices for totally symmetric arbitrary spin canonical and anomalous shadows were also obtained.

\noindent {\bf Light-cone gauge symmetries of anomalous currents and shadows}. To complete the light-cone gauge description of conformal currents  and shadows we have to work out realization of the $so(4,2)$ algebra symmetries on space of the ket-vectors $|\phi_\cur\rangle$, $|\phi_\sh\rangle$. Let us present transformation rules of currents and shadows under action of the $so(4,2)$ algebra transformations in the following way:
\be
\delta |\phi_\cur\rangle = G_\cur |\phi_\cur\rangle \,, \qquad
\delta |\phi_\sh\rangle = G_\sh |\phi_\sh\rangle,
\ee
where $G_\cur$ and $G_\sh$ stand for representation of generators of the $so(4,2)$ algebra in terms of differential operators acting on the respective light-cone gauge ket-vectors of currents and shadows. Expressions for $G_\cur$ and $G_\sh$ we found can be presented on an equal footing as
\beq
\label{15102014-01} && P^+=\partial^+\,,\quad P^i = \partial^i\,,\quad P^- = \partial^-\,,
\\
\label{15102014-02} && J^{+-} = x^+ \partial^-  - x^-\partial^+\,,
\\
\label{15102014-03} && J^{+i}=  x^+\partial^i - x^i\partial^+\,,
\\
\label{15102014-04} && J^{ij} = x^i\partial^j-x^j\partial^i + M^{ij}\,,
\\
\label{15102014-05} && J^{-i} = x^-\partial^i - x^i\partial^- + M^{-i}\,,
\\
\label{15102014-06} && D = x^+ \partial^- + x^-\partial^+ + x^i\partial^i+ \Delta\,,
\\
\label{15102014-07} && K^+ = K_\Delta^+\,,
\\
\label{15102014-08} && K^i = K_\Delta^i + M^{ij} x^j + \half\{ M^{i-},x^+\} + M^{\ominussm i}\,,
\\
\label{15102014-09} && K^- = K_\Delta^- + \half \{M^{-i}, x^i\}  - M^{\ominussm i}\frac{\partial^i}{\partial^+} + \frac{1}{\partial^+}B\,,
\\
\label{15102014-10} && \hspace{1cm}  K_\Delta^a \equiv - \half (2x^+x^-+x^jx^j) \partial^a + x^a D\,,\qquad a=\pm,i\,,
\\
\label{15102014-11} && \hspace{1cm} M^{-i} \equiv M^{ij} \frac{\partial^j}{\partial^+} + \frac{1}{\partial^+} M^{\oplussm i}\,, \qquad M^{i-} = -M^{-i}\,.
\eeq
For currents and shadows, operators $\Delta$, $M^{\Rsm\Lsm}$, $M^{\oplussm i}$, $M^{\ominussm i}$ appearing in \rf{15102014-01}-\rf{15102014-11} take the following form
\beq
\label{15102014-12} && \Delta_\cur = 2 + \nu\,, \qquad \nu = \kappa + S_1 - S_2\,,
\\
\label{15102014-14} && M_\cur^{\Rsm\Lsm} = S_1 + S_2\,,
\\
\label{15102014-15} && M_\cur^{\oplussm \Rsm} =  \Box g_1 S_1^\Rsm  + g_2 S_2^\Rsm \,,
\\
\label{15102014-16} && M_\cur^{\oplussm \Lsm}  =    S_1^\Lsm g_1 +  \Box S_2^\Lsm g_2 \,,
\\
\label{15102014-17} && M_\cur^{\ominussm \Rsm} = -(2\nu-1) g_1 S_1^\Rsm\,,
\\
\label{15102014-18} && M_\cur^{\ominussm \Lsm} = - S_2^\Lsm (2\nu+1)   g_2\,,
\\
\label{15102014-19} && B_\cur =  \kappa (S_1-S_2) + S_1^2 + S_2^2 - j_1(j_1+1) - j_2(j_2+1)\,,
\eeq

\beq
\label{15102014-20} && \Delta_\sh = 2 - \nu\,, \qquad \nu = \kappa + S_1 - S_2\,,
\\
\label{15102014-21} && M_\sh^{\Rsm\Lsm} = S_1 + S_2\,,
\\
\label{15102014-22} && M_\sh^{\oplussm \Rsm} =   g_1 S_1^\Rsm  + \Box g_2 S_2^\Rsm \,,
\\
\label{15102014-23} && M_\sh^{\oplussm \Lsm}  =   \Box  S_1^\Lsm g_1 +  S_2^\Lsm g_2 \,,
\\
\label{15102014-24} && M_\sh^{\ominussm \Rsm} =  (2\nu+1) g_2 S_2^\Rsm\,,
\\
\label{15102014-25} && M_\sh^{\ominussm \Lsm} =  S_1^\Lsm (2\nu-1)   g_1\,,
\\
\label{15102014-26} && B_\sh =  \kappa (S_1-S_2) + S_1^2 + S_2^2 - j_1(j_1+1) - j_2(j_2+1)\,,
\eeq
where expressions for $g_1$, $g_2$ are given in \rf{12102014-37}, \rf{12102014-38}, while the operators $S_\tau$ and $S_\tau^{\Rsm,\Lsm}$, $\tau=1,2$ are defined in \rf{12102014-09} and \rf{12102014-35}, \rf{12102014-36}. We note the following interesting commutator
\be \label{15102014-27}
[M^{\oplussm i},M^{\oplussm j}] = \Box M^{ij}\,.
\ee

\noindent {\bf Light-cone gauge symmetries of canonical mixed-symmetry (and self-dual) currents and shadows}. Relations in \rf{15102014-01}-\rf{15102014-26} provide light-cone gauge realization of the $so(4,2)$ algebra symmetries on space of mixed-symmetry anomalous currents and  shadows. From these expressions, we can easily obtain realization of the $so(4,2)$ algebra symmetries on space of mixed-symmetry canonical  currents and  shadows and self-dual currents and shadow. To get realization on space of the mixed-symmetry canonical currents and shadows, we use ket-vectors as in \rf{20102014-05} and set $\kappa=j_1+j_2$. To get realization on space of the self-dual currents and shadows we use ket-vectors as in \rf{20102014-07} and set $j_2=0$, $S_2=0$, $S_2^{\Rsm,\Lsm}=0$. Note that for the self-dual fields the quantity $g_1$ \rf{12102014-37} is simplified as $g_1=1$.

\newsection{AdS/CFT correspondence} \label{ads-cft}

We now study the AdS/CFT correspondence for mixed-symmetry AdS fields and
boundary mixed-symmetry conformal currents and shadows. To study the AdS/CFT
correspondence we use the light-cone gauge formulation for mixed-symmetry AdS fields, mixed-symmetry anomalous currents, and shadows  we developed in the preceding sections in this paper. We recall that our light-cone approach to mixed symmetry AdS fields leads to the decoupled equations of motion and surprisingly simple light-cone gauge Lagrangian. Owing this property of our approach a study of AdS/CFT correspondence for arbitrary spin mixed-symmetry AdS field becomes similar to the one for scalar AdS field.
We note also that the computation of action of mixed-symmetry AdS fields on solution of the Dirichlet problem is considerably simplified in our approach. Perhaps, this is the main advantage of our light-cone gauge approach.

In our study of AdS/CFT correspondence, we deal with light-cone gauge formulation not only at AdS side but also at the boundary CFT. This is to say that, in the
framework of our light-cone gauge formulation, the study of AdS/CFT correspondence implies the matching of:

\noindent \ibf) on-shell $so(4,2)$ light-cone gauge symmetries of bulk mixed-symmetry field and the corresponding $so(4,2)$ conformal light-cone gauge symmetries of mixed-symmetry boundary conformal current and shadow;

\noindent \iibf) Action of mixed-symmetry AdS field evaluated on the solution of
equations of motion with the Dirichlet problem corresponding
to the boundary mixed-symmetry shadow and the boundary two-point
 vertex for the shadow.

\subsection{ AdS/CFT for normalizable
modes of massive AdS field and anomalous conformal current}\label{ads-cft-current}

We now consider the AdS/CFT correspondence for mixed-symmetry massive AdS field and
mixed-symmetry anomalous conformal current. We begin with the discussion of the normalizable solution of equations of motion for massive AdS field.  To this end we note that equations of motion obtained from action in \rf{12102014-04}-\rf{12102014-07} takes the form
\be \label{14102014-03}
\bigl(\Box + \partial_z^2 -\frac{1}{z^2}(\nu^2 - \frac{1}{4})\bigr)|\phi\rangle = 0 \,.
\ee
The normalizable solution of Eq.\rf{14102014-03} is given by
\beq
\label{14102014-04} && |\phi(x,z)\rangle = U_\nu |\phi_\cur(x)\rangle \,,
\\
\label{14102014-05} && \hspace{1cm} U_\nu \equiv h_\kappa (-)^{S_2}
\sqrt{zq} J_\nu(zq) q^{-(\nu + \half)}\,,
\\
\label{14102014-06} && \hspace{1cm} h_\kappa\equiv 2^\kappa\Gamma(\kappa+1)\,,\qquad  q^2\equiv \Box\,, \qquad
\eeq
where $J_\nu$ in \rf{14102014-05} is the Bessel function. From \rf{14102014-04}, we see that solution for $\phik$ is governed by the operator $U_\nu$. We note the following relations which are helpful for the study of AdS/CFT correspondence:
\beq
\label{14102014-08} && \TT_{\nu-\half} U_\nu  = U_{\nu-1}\,,
\\
\label{14102014-09} && \TT_{-\nu-\half} U_\nu  = - U_{\nu+1}\Box\,,
\\
\label{14102014-10} && \TT_{\nu+\half} U_{\nu+1}  = U_\nu\,,
\\
\label{14102014-11} && \TT_{-\nu+\half} U_{\nu-1}  = - U_\nu \Box\,,
\\
\label{14102014-12} && z U_{\nu-1} + z \Box U_{\nu+1} =2\nu U_\nu\,,
\\
\label{14102014-14} && x^a U_\nu = U_\nu x^a + z U_{\nu+1} \partial^a\,,
\eeq
where $\TT_\nu$ is defined in \rf{12102014-38-x1}.
Relations \rf{14102014-08}-\rf{14102014-14} are valid on space of ket-vectors which depend on boundary coordinate $x^a$ and do not depend on the radial coordinate $z$. We note also that relations \rf{14102014-08}-\rf{14102014-14} can easily be proved by using the following well known formulas for the Bessel functions
\be
\TT_\nu J_{\nu } = J_{\nu-1}\,,  \qquad \TT_{-\nu} J_{\nu } = - J_{\nu + 1}\,.
\ee

The asymptotic behavior of solution \rf{14102014-04} takes the form
\be \label{14102014-07}
|\phi(x,z)\rangle \ \ \stackrel{z\rightarrow 0}{\longrightarrow} \ \ z^{\nu +
\half} \frac{2^\kappa\Gamma(\kappa+1)}{2^\nu\Gamma(\nu+1)}(-)^{S_2}|\phi_\cur(x)\rangle\,.
\ee
From \rf{14102014-07}, we see that, up to overall normalization factor, the ket-vector   $|\phi_\cur\rangle$ is indeed boundary value of the
normalizable solution.

{\bf Matching of bulk and boundary $so(4,2)$ symmetries}. In the framework of AdS/CFT correspondence, we expect that the boundary value $|\phi_\cur\rangle$ is realized as mixed-symmetry anomalous conformal current. To make sure that
$|\phi_\cur\rangle$ is indeed realized as mixed-symmetry anomalous conformal current we prove the following statement: {\it Bulk light-cone gauge $so(4,2)$ symmetries of the normalizable solution \rf{14102014-04} amount to boundary light-cone gauge $so(4,2)$ conformal symmetries of the anomalous conformal current}.

Thus our purpose is to demonstrate the matching of the $so(4,2)$ algebra generators for bulk mixed-symmetry massive AdS field given in \rf{12102014-18}-\rf{12102014-27} and the ones for the boundary mixed-symmetry anomalous conformal current given in  \rf{15102014-01}-\rf{15102014-09} and \rf{15102014-12}-\rf{15102014-19}. To this end we note that all that is required is to match the following generators of the $so(4,2)$ algebra
\be \label{15102014-28}
P^\pm \,, \qquad P^i\,, \qquad J^{+-}\,, \qquad J^{\pm i}\,, \qquad J^{ij}\,, \qquad D\,, \qquad K^+\,.
\ee
This is to say that if generators \rf{15102014-28} match then the remaining generators $K^-$ and $K^i$ match automatically in view of the commutation relations of the $so(4,2)$ algebra.

Using notation $G_\AdSsm$ and $G_\CFTsm$ for the respective bulk and boundary generators given  in \rf{12102014-18}-\rf{12102014-27} and \rf{15102014-01}-\rf{15102014-09}, we note that the matching implies that we should verify the following relation for the solution in \rf{14102014-04} and for the bulk and boundary generators
\be \label{15102014-29}
G_\AdSsm \phik = U_\nu G_\CFTsm |\phi_\cur\rangle\,.
\ee
We now verify relation \rf{15102014-29} for generators of the $so(4,2)$ algebra given in \rf{15102014-28} in turn.

\noindent \ibf) We start with matching the following generators:
\be \label{15102014-30}
P^+\,, \qquad P^i\,, \qquad J^{+i}\,, \qquad J^{ij}\,.
\ee
From \rf{12102014-18}-\rf{12102014-21} and \rf{15102014-01}-\rf{15102014-04}, we see that bulk and boundary generators given in \rf{15102014-30} coincide. As generators \rf{15102014-30} commute with the operator $U_\nu$ \rf{14102014-05}, we conclude that generators \rf{15102014-30} satisfy relation \rf{15102014-29}.

\noindent \iibf) We now match generators $P_\AdSsm^-$ and $P_\CFTsm^-$ given in \rf{12102014-25} and \rf{15102014-01} respectively. To this end we note that operator $\MM^2$ \rf{12102014-27-a} entering $P_\AdSsm^-$ \rf{12102014-25} can be represented in terms of operator $\TT_\nu$ \rf{12102014-38-x1} as
\be \label{15102014-31}
\MM^2 = - \TT_{-\nu+\half} \TT_{\nu-\half}\,.
\ee
Using \rf{14102014-08},\rf{14102014-09} and \rf{15102014-31}, we find the following remarkable relation
\be \label{15102014-32}
\MM^2 U_\nu =  U_\nu \Box\,.
\ee
Relation \rf{15102014-32} is valid on space of ket-vectors which are independent of the radial coordinate $z$. Taking into account \rf{12102014-06}, we note the simple relation
\be \label{15102014-33}
-\frac{\partial^i\partial^i}{2\partial^+} + \frac{1}{2\partial^+} \Box = \partial^-\,.
\ee
From \rf{15102014-32},\rf{15102014-33}, we see that $P_\AdSsm^-$ and $P_\CFTsm^-$ satisfy the relation \rf{15102014-29}. Note that, because operator $\partial^-$ commutes with $U_\nu$, relation \rf{15102014-29} implies that on the space of solution \rf{14102014-04} one has the relation
\be \label{15102014-34}
P_\AdSsm^- \approx \partial^-\,.
\ee
In \rf{15102014-34} and below, the notation $\approx$ is used to indicate the fact that relation \rf{15102014-34} is valid on the space of solution $\phik$ in \rf{14102014-04}.

\noindent \iiibf) Next we match the generators $J_\AdSsm^{+-}$, $D_\AdSsm$ \rf{12102014-19} \rf{12102014-22} and $J_\CFTsm^{+-}$, $D_\CFTsm$ \rf{15102014-02}, \rf{15102014-06}. To math these generators we use relation \rf{15102014-34} which implies that on space of solution \rf{14102014-04} expressions for $J_\AdSsm^{+-}$, $D_\AdSsm$  can be represented as
\beq
\label{15102014-35} && J_\AdSsm^{+-} \approx x^+ \partial^- - x^- \partial^+\,,
\\
\label{15102014-36} && D_\AdSsm \approx x^+ \partial^- + x^- \partial^+ + x^i \partial^i + z\partial_z + \frac{3}{2}\,.
\eeq
From \rf{15102014-35} and \rf{15102014-02}, we see that generators $J_\AdSsm^{+-}$ and  $J_\CFTsm^{+-}$ coincide. As these generators commute with operator $U_\nu$ \rf{14102014-05}, we conclude that generators $J_\AdSsm^{+-}$, $J_\CFTsm^{+-}$ satisfy relation \rf{15102014-29}. Using \rf{15102014-36} and \rf{15102014-06},\rf{15102014-12}, we verify that $D_\AdSsm$ and $D_\CFTsm$ also satisfy relation \rf{15102014-29}.

\noindent \ivbf) We proceed with matching  of generators $J_\AdSsm^{-i}$ \rf{12102014-26}, $J_\CFTsm^{-i}$ \rf{15102014-05}.
To match these generators we use relation \rf{15102014-34} which implies that on space of solution \rf{14102014-04} expressions for $J_\AdSsm^{-i}$  can be represented as
\be \label{15102014-37}
J_\AdSsm^{-i} \approx x^- \partial^i - x^i\partial^- + M_\AdSsm^{-i}\,,
\ee
where $ M_\AdSsm^{-i}$ is given in \rf{12102014-28},\rf{12102014-33},\rf{12102014-34}. Comparing \rf{15102014-37} with \rf{15102014-05},\rf{15102014-11}, we see that all that remains is to match $M_\AdSsm^{\oplussm i}$ in \rf{12102014-33}, \rf{12102014-34} and $M_\CFTsm^{\oplussm i}$ in \rf{15102014-15},\rf{15102014-16}. In other words, we should prove the relation
\be \label{15102014-38}
M_\AdSsm^{\oplussm i}\phik = U_\nu M_\CFTsm^{\oplussm i} |\phi_\cur\rangle\,.
\ee
Using \rf{14102014-08}-\rf{14102014-11}, we verify that relation \rf{15102014-38} holds true.

\noindent \vbf) Finally, we match generators $K_\AdSsm^+$ \rf{12102014-23}, $K_\CFTsm^+$ \rf{15102014-07}. To this end we note the following useful relation
\beq
\label{16102014-01} && \bigl(K_\Delta^a - \half z^2 \partial^a\bigr)\Bigr|_{ \Delta= z\partial_z+ \frac{3}{2} }  U_\nu = U_\nu K_\Delta^a\Bigr|_{ \Delta= 2 + \nu}\,,
\\
&& K_\Delta^a \equiv -\half x^2 \partial^a + x^a (x\partial + \Delta)\,, \qquad a=\pm,i\,,
\eeq
where $x^2\equiv x^a x^a$, $x \partial \equiv x^a\partial^a$. Relation \rf{16102014-01} can straightforwardly be proved by using \rf{14102014-12}, \rf{14102014-14}. Using \rf{16102014-01} for $a=+$, we see that $K_\AdSsm^+$ \rf{12102014-23} and $K_\CFTsm^+$ \rf{15102014-07} satisfy relation \rf{15102014-29}.
Thus the generators $K_\AdSsm^+$ and $K_\CFTsm^+$ match. As a side remark we note that, by using \rf{16102014-01}, we checked explicitly that the remaining generators $K^i$, $K^-$ also satisfy relation  \rf{15102014-29}.

\subsection{ AdS/CFT for non-normalizable
modes of massive AdS field and anomalous shadow }

In this Section, we consider the AdS/CFT correspondence for mixed-symmetry massive AdS field and mixed-symmetry anomalous shadow. We begin with the discussion of the non-normalizable solution of equations of motion for massive AdS field.  To this end we note that non-normalizable solution of equations of motion \rf{14102014-03} with the
Dirichlet problem corresponding to anomalous shadow takes the form
\beq
\label{16102014-02} |\phi(x,z)\rangle  & = &  \sigma_\nu \int d^4 y\, G_\nu (x-y,z)
|\phi_\sh(y)\rangle\,,
\\
\label{16102014-03} && G_\nu(x,z) = \frac{c_\nu z^{\nu+\half}}{ (z^2+
|x|^2)^{\nu + 2} }\,,
\\
\label{16102014-04} &&  c_\nu \equiv \frac{\Gamma(\nu+2)}{\pi^2 \Gamma(\nu)} \,,
\\
\label{16102014-05} && \sigma_\nu \equiv \frac{2^\nu\Gamma(\nu)}{
2^\kappa\Gamma(\kappa)}(-)^{S_2} \,,
\eeq
where $G_\nu$ in \rf{16102014-03} is the Green function. From \rf{16102014-02}, we see that solution for $\phik$ is governed by the $\sigma_\nu G_\nu$.

The asymptotic behavior of solution \rf{16102014-02}  takes the form
\be \label{16102014-12}
|\phi(x,z)\rangle  \,\,\, \stackrel{z\rightarrow 0 }{\longrightarrow}\,\,\,
z^{-\nu + \half} \sigma_\nu |\phi_\sh(x)\rangle\,.
\ee
From \rf{16102014-12}, we see that, up to overall normalization factor, the ket-vector   $|\phi_\sh\rangle$ is indeed boundary value of the non-normalizable solution.

In the framework of AdS/CFT correspondence, we expect that the boundary value $|\phi_\sh\rangle$ is realized as mixed-symmetry anomalous shadow. To make sure that $|\phi_\sh\rangle$ is indeed realized as mixed-symmetry anomalous shadow we prove the following two statements:

\noindent \ibf) {\it Bulk light-cone gauge $so(4,2)$ symmetries of the non-normalizable solution \rf{16102014-02} amount to boundary light-cone gauge $so(4,2)$ conformal symmetries of the mixed-symmetry anomalous shadow}.

\noindent \iibf) {\it action of mixed-symmetry AdS field evaluated on solution \rf{16102014-02}  coincides, up to normalization factor, with boundary 2-point light-cone gauge vertex for the mixed-symmetry
anomalous shadow}.

We now prove these statements in turn.

\noindent {\bf Matching of bulk and boundary $so(4,2)$ symmetries}. Our purpose is to demonstrate the matching of the $so(4,2)$ algebra generators for bulk mixed-symmetry massive AdS field given in \rf{12102014-18}-\rf{12102014-27} and the ones for the boundary mixed-symmetry anomalous shadow given in  \rf{15102014-01}-\rf{15102014-09} and \rf{15102014-20}-\rf{15102014-26}.
Let us use the notation $G_\AdSsm$ and $G_\CFTsm$ for the respective bulk and boundary generators given  in \rf{12102014-18}-\rf{12102014-27} and \rf{15102014-01}-\rf{15102014-09}. All that is required then is to verify that these generators satisfy the following relation
\be \label{16102014-14}
G_\AdSsm  |\phi(x,z)\rangle   =   \sigma_\nu \int d^4 y\, G_\nu (x-y,z)
G_\CFTsm |\phi_\sh(y)\rangle\,,
\ee
Procedure for checking relation \rf{16102014-14} is the same as the one we used for the case of conformal current in Sec.\ref{ads-cft-current}. Therefore to avoid the repetition we just present the following relations which are helpful in the studying of AdS/CFT correspondence for the case of shadow:
\beq
\label{16102014-06} && \TT_{\nu-\half}(\sigma_\nu G_\nu ) =  \Box \sigma_{\nu-1} G_{\nu-1}\,,
\\
\label{16102014-07} && \TT_{-\nu - \half}(\sigma_\nu G_\nu ) = - \sigma_{\nu+1} G_{\nu+1}\,,
\\
\label{16102014-08} && \TT_{\nu+\half}(\sigma_{\nu+1} G_{\nu+1}) =  \Box \sigma_\nu G_\nu\,,
\\
\label{16102014-09} && \TT_{-\nu+\half}(\sigma_{\nu-1} G_{\nu-1}) = - \sigma_\nu G_\nu\,,
\\
\label{16102014-10} && z \Box \sigma_{\nu-1} G_{\nu-1} + z \sigma_{\nu+1} G_{\nu+1} = 2\nu G_\nu\,,
\\
\label{16102014-11} && x^a \sigma_\nu G_\nu = \sigma_\nu G_\nu y^a - z \partial^a ( \sigma_{\nu-1} G_{\nu-1})\,,
\eeq
where the operator $\TT_\nu$ is defined in \rf{12102014-38-x1}.

\noindent {\bf Effective action for mixed-symmetry anomalous shadow}. In this paper, action of AdS field  evaluated on non-normalizable solution in \rf{16102014-02} is referred to as effective action.
In order to find the effective action, we should plug the solution given in \rf{16102014-02} into light-cone gauge action \rf{12102014-04} with Lagrangian given in \rf{12102014-05}. Note however that we should add to the light-cone gauge action an appropriate boundary term. Using the method for finding boundary term developed in Ref.\cite{Arutyunov:1998ve}, we verify that a light-cone gauge Lagrangian which involves an appropriate boundary term can be presented in the following way
\be \label{16102014-15}
\LL  =    \langle \partial^a \phi|   | \partial^a \phi\rangle +
\langle \TT_{\nu-\half} \phi| | \TT_{\nu-\half} \phi\rangle\,,
\ee
Also note that, to adapt our expressions to commonly used Euclidean signature, we have changed sign of Lagrangian, $\LL\rightarrow - \LL$, when passing from \rf{12102014-05} to \rf{16102014-15}.
It is easy to see that light-cone gauge action \rf{12102014-04}, \rf{16102014-15} considered on solution of equations of motion can be represented in the following way
\be  \label{16102014-16}
- S_\eff  =  \int d^4 x\,  \LL_\eff\Bigr|_{z\rightarrow
0} \,, \qquad  \ \ \
\LL_\eff \equiv   \phibr \TT_{\nu -\half } \phik \,.
\ee
Plugging solution \rf{16102014-02} into  \rf{16102014-16}, we get the following effective action for mixed-symmetry anomalous shadow
\be \label{16102014-17}
-S_\eff  =   2\kappa c_\kappa \Gamma^{\rm sh-sh} \,,
\ee
where $\Gamma^{\rm sh-sh}$ takes the same form as in \rf{16102014-19}, while $c_\kappa$ is defined in \rf{16102014-04}. Thus we see that, up to overall normalization factor given by $2\kappa c_\kappa$, the action of the mixed-symmetry massive AdS field evaluated on the non-normalizable solution amounts to the 2-point vertex of mixed-symmetry anomalous shadow. Our analysis allows us to find the normalization factor which might be important for the systematical study of the AdS/CFT correspondence.

\noindent {\bf Effective actions for canonical mixed-symmetry shadow, self-dual shadow, and canonical totally symmetric shadow}. Expression for $S_\eff$ in \rf{16102014-17} gives effective action for mixed-symmetry anomalous shadow. From this expression, we can easily obtain effective action for mixed-symmetry canonical shadow, self-dual shadow, and canonical totally symmetric shadow. To get $S_\eff$ for canonical mixed-symmetry shadow we use $\Gamma^{\sh-\sh}$ given in  \rf{16102014-19}, \rf{20102014-06} and set $\kappa=j_1+j_2$ in \rf{16102014-17}. $S_\eff$ for canonical totally symmetric shadow is obtained by using $\Gamma^{\sh-\sh}$ given in  \rf{16102014-19}, \rf{20102014-14-a1}  and setting $\kappa= 2j_1$, $j_1=j$, $j_2=j$ in \rf{16102014-17}. To get $S_\eff$  for the self-dual shadow we use $\Gamma^{\sh-\sh}$ given in \rf{16102014-19}, \rf{20102014-08}.

The following remarks are in order.

\noindent \ibf) In the Lorentz covariant framework, effective action for mixed-symmetry canonical shadow in $R^{d-1,1}$ with arbitrary $d$ and particular values  $j_1=3/2$, $j_2=1/2$ was found in Ref.\cite{Alkalaev:2012ic}.

\noindent \iibf) In Lorentz covariant framework, effective action for self-dual shadow in $R^{3,1}$ with arbitrary $\kappa$ and the particular value of the label $j_1=1$ was discussed in Ref.\cite{Arutyunov:1998xt}.

\noindent \iiibf) The intertwining operator realization of AdS/CFT correspondence was studied in Refs.\cite{Dobrev:1998md}.

In conclusion of this section, we note that we applied our light-cone gauge formulation for the study of the AdS/CFT correspondence for mixed-symmetry AdS fields and boundary mixed-symmetry currents and shadows. Beyond this, we expect that our light-cone gauge approach might have other interesting applications to studying AdS/CFT correspondence along the lines in Refs.\cite{Koch:2010cy}-\cite{Ananth:2012tf}.

\newsection{ Mixed-symmetry conformal fields in $R^{3,1}$}\label{conf-field}

For shadows in $R^{3,1}$, a kernel of the effective action in \rf{16102014-19-a1} is not well defined when $\kappa$ \rf{13102014-18-x1} takes integer values (see, e.g., Ref.\cite{Aref'eva:1998nn}). We note however that the kernel becomes well defined for shadows having non-integer $\kappa$. Taking this into account, we use non-integer $\kappa$ to regularize the kernel of effective action. When removing the regularization, we are left with a logarithmic divergence of the effective action. This logarithmic divergence of the effective action for shadow turns out to be light-cone gauge action of mixed-symmetry conformal field.

We now explain the regularization procedure. Using the notation $\kappa_\intrm $ for integer part of $\kappa$, we introduce the regularization parameter $\varepsilon$ by the relation
\be  \label{16102014-20}
\kappa - \kappa_\intrm = - 2\varepsilon\,,\qquad  \qquad \kappa_\intrm -\hbox{ integer}.
\ee
Using  \rf{16102014-20} and taking into account the $\nu$ in \rf{12102014-08}, we note the following textbook asymptotic behavior for the kernel:
\beq
\label{16102014-21} && \frac{1}{|x|^{2\nu + 4}}\,\,\, \stackrel{\varepsilon \sim
0}{\mbox{\Large$\sim$}}\,\,\, \frac{1}{\varepsilon} \varrho_{\nu_\intrm} \Box^{\nu_{\intrm}}
\delta^{(4)}(x)\,, \qquad   \varrho_\nu \equiv  \frac{\pi^2}{4^\nu \Gamma(\nu
+ 1)\Gamma(\nu + 2)}\,,
\\
\label{16102014-21-a1}&&  \nu_\intrm \equiv \kappa_\intrm + S_1  - S_2 \,.
\eeq
Plugging \rf{16102014-21} into expression for $\Gamma^{\rm sh-sh}$ in \rf{16102014-19}, we obtain
\beq
&& \Gamma\,\,\, \stackrel{ \varepsilon \sim
0}{\mbox{\Large$\sim$}}\,\,\, \frac{1}{\varepsilon} \varrho_{\kappa_\intrm} \int
d^4 x\,\, \LL\,, \hspace{1cm}
\nonumber\\
\label{16102014-22} && \LL =   \phibr    \Box^{\nu_\intrm} \phik \,.
\eeq
In \rf{16102014-22}, to simplify the notation, we use  the  identifications of the ket-vectors, $\phik\equiv |\phi_\sh\rangle$, when passing from  \rf{16102014-19} to \rf{16102014-22}. Lagrangian \rf{16102014-22} provides light-cone gauge description of mixed-symmetry conformal field. Taking into account values of $\kappa$ given in \rf{13102014-18-x1}, \rf{13102014-15}-\rf{13102014-17}, we consider three cases of mixed-symmetry conformal fields
\beq
\label{16102014-23} &&  \hspace{-2cm} \kappa_\intrm = j_1 + j_2 + N, \quad j_1 j_2 = 0, \quad N = 0,1,\ldots , \hspace{1.1cm} \hbox{self-dual conformal fields}
\\
\label{16102014-24} && \hspace{-2cm} \kappa_\intrm = j_1 + j_2, \hspace{1.3cm}  j_1 j_2 \ne 0, \hspace{4cm} \hbox{short mix-sym. conformal fields}
\\
\label{16102014-25}&&  \hspace{-2cm} \kappa_\intrm = j_1 + j_2 + N, \quad j_1j_2 \ne 0,\quad  N =1,2,\ldots, \hspace{1.1cm}  \hbox{long mix-sym. conformal fields}
\eeq
We recall that conformal dimensions of fields \rf{16102014-23}-\rf{16102014-25} are given by
\be \label{17102014-11}
\Delta = 2 - \kappa_\intrm \,.
\ee

\noindent {\bf Long mixed-symmetry conformal field.}%
\footnote{ In this paper, terminology long conformal field is used to indicate the fact that, in the framework of AdS/CFT correspondence, our long conformal fields are dual to massive AdS fields with integer values of $E_0$ defined by the relations \rf{13102014-18-x1}, \rf{16102014-25}. Note also that, in the framework of AdS/CFT correspondence, our short conformal fields are dual to massless AdS fields.}
In terms of component fields, Lagrangian for long mixed-symmetry conformal field is obtained by using relations \rf{16102014-21-a1}, \rf{16102014-25} and by plugging a ket-vector $\phik$, which takes the same form as in \rf{12102014-02}, into \rf{16102014-22}. Doing so, we get
\be \label{17102014-12}
\LL  =   \sum_{m_1=-j_1}^{j_1} \sum_{m_2=-j_2}^{j_2} \phi_{m_1,m_2}^\dagger \Box^{j_1  + j_2 + N+  m_1-m_2} \phi_{m_1,m_2}\,, \qquad N=1,2,\ldots \,.
\ee
As all fields in \rf{17102014-12} are complex-valued it is easy to see that number of {\it complex-valued} on-shell D.o.F described by Lagrangian \rf{17102014-12} is given by
\be \label{17102014-14}
N_{DoF}^\Co = (2j_1+1)(2j_2+1)(j_1 + j_2 + N)\,.
\ee
Lagrangian \rf{17102014-12} describes conformal fields associated with direct sum of two non-unitary representations of the $so(4,2)$ algebra labelled by  $\Delta,j_1,j_2$ and  $\Delta,j_2,j_1$, where $\Delta$ is given in \rf{17102014-11},\rf{16102014-25}. For totally symmetric conformal fields, $j_1 = j_2$, we can impose the reality condition on the space of fields, $\phi_{-m_2,-m_1}^\dagger = \phi_{m_1,m_2}$.

\noindent {\bf Short mixed-symmetry conformal field}. In terms of component fields, Lagrangian for short mixed-symmetry conformal field is obtained by using relations \rf{16102014-21-a1}, \rf{16102014-24} and by plugging a ket-vector $\phik$, which takes the same form as in \rf{13102014-07}, into \rf{16102014-22}. Doing so, we get
\beq
\label{17102014-04} && \hspace{-2.3cm} \LL = \LL_0 + \LL_1 + \LL_2 \,,
\\
\label{17102014-05} \LL_0  &  = & \phi_{-j_1,j_2}^\dagger \phi_{-j_1,j_2}\,,
\\
\label{17102014-06} \LL_1 &= & \sum_{m=-j_1+1}^{j_1} \phi_{m,j_2}^\dagger \Box^{ j_1+m} \phi_{m,j_2}\,,
\\
\label{17102014-07} \LL_2 &= & \sum_{m=-j_2+1}^{j_2} \phi_{-j_1,-m}^\dagger \Box^{j_2+m} \phi_{-j_1,-m}\,.
\eeq
From \rf{17102014-05}, we see that the field $\phi_{-j_1,j_2}$ is equal to zero on-shell. In other words, the field $\phi_{-j_1,j_2}$ does not describe propagating degrees of freedom. Therefore propagating D.o.F of short mixed-symmetry conformal field are described by the Lagrangian
\be  \label{17102014-08}
\LL  =  \LL_1 + \LL_2 \,.
\ee
It is easy to see that, a number of {\it complex-valued} on-shell D.o.F described by Lagrangian \rf{17102014-08}, \rf{17102014-06}, \rf{17102014-07} is given by
\be \label{17102014-09}
N_{\rm D.o.F}^\Co = j_1(2j_1+1) + j_2(2j_2+1)\,.
\ee
Lagrangian \rf{17102014-08} describes conformal fields associated with direct sum of two non-unitary representations of the $so(4,2)$ algebra  labelled by $\Delta,j_1,j_2$ and $\Delta,j_2,j_1$, where $\Delta$ is given in \rf{17102014-11},\rf{16102014-24}. For the discussion of short totally symmetric conformal fields, $j_1 = j_2$, see below.

\noindent {\bf Self-dual conformal field}. For the case of self-dual conformal fields \rf{16102014-23}, one of the labels $j_1$, $j_2$ should be equal to zero. Let us set $j_2=0$. Then, in terms of component fields, Lagrangian for self-dual conformal field is obtained by using relations \rf{16102014-21-a1}, \rf{16102014-23} and plugging a ket-vector $\phik$, which takes the same form as in \rf{13102014-22}, into \rf{16102014-22}. Doing so, we get
\be \label{17102014-15}
\LL=  \sum_{m = -j_1}^{j_1} \phi_{m,0}^\dagger \Box^{j_1 + N + m}\phi_{m,0}\,, \qquad  N = 0, 1,2,\ldots\,.
\ee
Number of {\it complex-valued} on-shell D.o.F described by Lagrangian \rf{17102014-15} is given by
\be \label{17102014-16}
N_{\rm D.o.F}^\Co = (2j_1+1)(j_1  + N)\,.
\ee
Lagrangian \rf{17102014-15} describes conformal fields associated with direct sum of two non-unitary representation of the $so(4,2)$ algebra  labelled by  $\Delta,j_1,0$ and $\Delta,0,j_1$, where $\Delta$ is given in \rf{17102014-11},\rf{16102014-23}.

Simplest self-dual conformal field described by Lagrangian \rf{17102014-15} corresponds to the case $j_1=1$, $N=0$. Such self-dual conformal field appears in the field content of ${\cal N}=4$, $4d$ conformal supergravity (for review, see Ref.\cite{Fradkin:1985am}).
Also we note that, for the self-dual conformal field with $j_1=1$, $N=0$, ordinary-derivative light-cone gauge Lagrangian was obtained in Ref.\cite{Metsaev:2008ba}.

\noindent {\bf Short totally symmetric conformal field}. For totally symmetric bosonic fields, we have the condition $j_1 = j_2$. Using the notation $j\equiv j_1$, we can impose the reality condition
\be
\label{22102014-09}  \phi_{m,j}^\dagger = \phi_{-j,-m}\,, \qquad m= -j+1, -j+2, \ldots , j\,.
\ee
Introducing then  a set of complex-valued fields $\phi_m$ by the relations
\be \label{22102014-10}
\phi_{j+m} \equiv \sqrt{2} \phi_{m,j}\,, \qquad m =-j+1,-j+2,\ldots j\,,
\ee
we see that Lagrangian \rf{17102014-08},\rf{17102014-06},\rf{17102014-07} takes the form
\be\label{22102014-11}
\LL = \sum_{m=1}^{2j} \phi_m^\dagger \Box^m \phi_m\,.
\ee
Using the notation $s\equiv 2j$, we note that Lagrangian \rf{22102014-11} gives light-cone gauge description of totally symmetric spin-$s$ conformal field which, in the framework of Lorentz covariant approach, was discussed for first in Ref.\cite{Fradkin:1985am}. From \rf{22102014-11}, it is easy to see that number of real-valued on-shell D.o.F. described by Lagrangian \rf{20102014-02} is equal to $s(s+1)$. Thus we see that our light-cone gauge Lagrangian leads to the number of D.o.F found in Ref.\cite{Fradkin:1985am}.

Let us conclude with following remarks.

\noindent \ibf) All conformal unitary representations of the $so(d+1,2)$ algebra which are associated with fields in $R^{d,1}$ and $AdS_{d+1}$ were classified in Refs.\cite{Siegel:1988gd,Metsaev:1995jp}. Our Lagrangian of conformal fields involves higher-derivatives. From the expressions for Lagrangian given in \rf{17102014-12}, \rf{17102014-04}, \rf{17102014-15} is is clear that our conformal fields are related to non-unitary representation of the $so(4,2)$ algebra. This is to say that conformal fields considered in this paper should not be confused with the ones studied in Refs.\cite{Siegel:1988gd,Metsaev:1995jp}.

\noindent \iibf) In the framework of Lorentz covariant formulation, a wide class of mixed-symmetry conformal fields was studied in Refs.\cite{Vasiliev:2009ck,Marnelius:2009uw}. We expect that our short mixed-symmetry conformal fields might be related to the ones studied in Refs.\cite{Vasiliev:2009ck,Marnelius:2009uw}. We note however that precise relation of our mixed-symmetry conformal fields to the ones in Refs.\cite{Vasiliev:2009ck,Marnelius:2009uw} is still to be understood.

\noindent \iiibf) Our study leads to simple expressions for the light-cone gauge Lagrangian of conformal fields. Therefore we believe that our light-cone gauge formulation of conformal fields might provide us new interesting possibilities for the studying partition functions of conformal fields along the lines in Refs.\cite{Tseytlin:2013jya}. In the framework of BRST approach, discussion of partition functions of conformal fields may be found in Refs.\cite{Metsaev:2014iwa,Metsaev:2014vda}.

\noindent \ivbf) In the context of AdS/CFT correspondence, scalar long conformal fields were discussed in Ref.\cite{Diaz:2008hy}.

\noindent \vbf) In the framework AdS/CFT correspondence, short conformal fields are dual to massless AdS fields, while long conformal fields are dual to massive AdS fields having  integer values of lowest eigenvalue of the energy operator $E_0$. Taking this into account, it is difficult to overcome a desire to speculate on some special regime in the type IIB superstring theory in $AdS_5 \times S^5$ background when $E_0$ for all massive higher-spin fields take integer values. Such conjectured regime in the superstring theory should lead via AdS/CFT to stringy theory of conformal fields which  involves low-spin short conformal fields and higher-spin long conformal fields.

\noindent

\bigskip

{\bf Acknowledgments}. This work was supported by the Russian Science Foundation grant 14-42-00047.

\setcounter{section}{0} \setcounter{subsection}{0}

\appendix{Notation and conventions}

We use the Poincar\'e parametrization of $AdS_5$ space with a line element given by
\be \label{22102014-01}
ds^2= \frac{1}{z^2}(- dx^0 dx^0 + dx^idx^i + dx^3 dx^3
+ dz dz)\,,  \qquad i=1,2\,.
\ee
The light-cone coordinates $x^\pm$ and complex coordinates  $x^\Rsm$, $x^\Lsm$ are defined by the relations
\beq
&& x^\pm \equiv  \frac{1}{\sqrt{2}} (x^3 \pm x^0)\,,
\\
\label{22102014-03}
&& x^\Rsm \equiv \frac{1}{\sqrt{2}} (x^1 + \irm x^2)\,, \qquad
x^\Lsm \equiv \frac{1}{\sqrt{2}} (x^1 - \irm x^2)\,.
\eeq
The coordinate $x^+$ is considered as the light-cone evolution parameter. For derivatives with respect to $x^i$, $i=1,2$, we use the notation $\partial^i=\partial_i \equiv \partial/\partial x^i$. Also, for derivatives, we adopt the following conventions:
\be \label{22102014-02}
\partial^\pm=\partial_\mp \equiv \partial/\partial x^\mp\,, \qquad \partial^\Rsm = \partial_\Lsm \equiv \partial/\partial x^\Lsm\,,\qquad \partial^\Lsm = \partial_\Rsm \equiv \partial/\partial x^\Rsm\,, \qquad \partial_z =\partial / \partial z \,,
\ee
and the D'Alembertian operator $\Box$ in $R^{3,1}$ takes the form $\Box = 2\partial^+\partial^- + 2 \partial^\Rsm \partial^\Lsm$. In frame of the coordinates $x^\pm$, $x^\Rsm$, $x^\Lsm$, a product of two $so(3,1)$ algebra vectors $X^a$, $Y^a$ is decomposed as
\beq
\label{22102014-04} X^a Y^a & = &  X^+ Y^- + X^- Y^+ + X^i Y^i \,,
\\
\label{22102014-05} && X^i Y^i = X^\Rsm Y^\Lsm + X^\Lsm Y^\Rsm\,,
\eeq
where $X^a Y^a\equiv \eta_{ab} X^a Y^b$ and $\eta_{ab}$ stands for the flat metric $\eta_{ab} = (-,+++)$.

$so(4,2)$ algebra consists of translation generators $P^a$, conformal boost generators $K^a$, dilatation generator $D$, and generators of the Lorentz $so(3,1)$ algebra $J^{ab}$. We use the following commutations relations of the $so(4,2)$ algebra
\beq
&& {}[D,P^a]=-P^a\,, \hspace{2.5cm}  [P^a,J^{bc}]=\eta^{ab}P^c -\eta^{ac}P^b
\,,
\nonumber\\
\label{22102014-06} && [D,K^a]=K^a\,, \hspace{2.7cm} [K^a,J^{bc}]=\eta^{ab}K^c - \eta^{ac}K^b\,,
\\
&& [P^a,K^b]=\eta^{ab}D - J^{ab}\,, \hspace{1.2cm}  [J^{ab},J^{ce}]=\eta^{bc}J^{ae}+3\hbox{ terms} \,,
\nonumber
\eeq
In frame of the coordinates $x^\pm$, $x^\Rsm$, $x^\Lsm$, the generators are decomposed as
\beq
\label{22102014-07} && P^a = P^\pm, \ P^\Rsm\,, \ P^\Lsm\,, \quad K^a = K^\pm\,, K^\Rsm\,, K^\Lsm\,,
\\
\label{22102014-08} && J^{ab} = J^{+-}\,, \ J^{\pm \Rsm}\,, \ J^{\pm \Lsm}\,, \ J^{\Rsm\Lsm}\,.
\eeq
Commutators for generators in \rf{22102014-07},\rf{22102014-08} are obtained from the ones in \rf{22102014-06} by noticing that, in frame of  the coordinates $x^\pm$, $x^\Rsm$, $x^\Lsm$, the non-zero values of the flat metric $\eta^{ab}$ are given by  $\eta^{+-}=1$, $\eta^{\Rsm\Lsm} =1$. We adopt the following hermitian conjugation rules for the generators of the $so(4,2)$ algebra
\beq
&& P^{\pm \dagger} = - P^\pm\,, \quad P^{\Rsm \dagger } = -P^\Lsm\,,
\quad K^{\pm \dagger} = - K^\pm\,, \quad K^{\Rsm \dagger } = - K^\Lsm\,,
\\
&& J^{+-\dagger} = -J^{+-}\,, \quad J^{\pm \Rsm\dagger} = - J^{\pm \Lsm}\,, \quad J^{\Rsm\Lsm\dagger} = J^{\Rsm\Lsm}\,, \quad D^\dagger = - D\,.
\eeq

\end{document}